\documentclass[aps,preprint]{revtex4}
\usepackage{times}
\usepackage{graphicx}

\begin{document}

\title{Knotting of random ring polymers in confined spaces}

\author{C. Micheletti$^1$, D. Marenduzzo$^2$, E. Orlandini$^3$,
D. W. Sumners$^4$}

\affiliation{$^1$ International School for Advanced Studies (SISSA)
  and INFM, via Beirut 2-4, 34100 Trieste, Italy \\
$^2$ Mathematics Institute, University of Warwick, Coventry, CV4 7AL, UK\\
$^3$ Dipartimento di Fisica and
Sezione INFN, Universit\`a di Padova,
Via Marzolo 8, 35121 Padova, Italy \\
$^4$ Department of Mathematics, Florida State University,
Tallahasse, FL 32306, USA}

\begin{abstract}
Stochastic simulations are used to characterize the knotting
distributions of random ring polymers confined in spheres of various
radii. The approach is based on the use of multiple Markov chains and
reweighting techniques, combined with effective strategies for
simplifying the geometrical complexity of ring conformations without
altering their knot type. By these means we extend previous studies
and characterize in detail how the probability to form a given prime
or composite knot behaves in terms of the number of ring segments,
$N$, and confining radius, $R$. For $ 50 \le N \le 450 $ we show that
the probability of forming a composite knot rises significantly with
the confinement, while the occurrence probability of prime knots are,
in general, non-monotonic functions of $1/R$.  The dependence of other
geometrical indicators, such as writhe and chirality, in terms of $R$
and $N$ is also characterized. It is found that the writhe
distribution broadens as the confining sphere narrows.
\end{abstract}

\maketitle

\section{Introduction}

In recent years, novel motivations to characterize the properties of
knotted ring polymers have been provided by {\em in vivo} and {\em in
vitro} experiments on molecules of biological interest. Quantitative
estimates for the occurrence of various types of knots are
particularly abundant for the case of circular DNA, which can be
manipulated and probed by a variety of physico-chemical techniques,
such as gel electrophoresis and others (see
e.g. Ref. \cite{stasiak_gel,cozzarelli,stasiak_gel2}).  The available
data provides crucial benchmarks for theoretical models aimed at
describing the occurrence of knots in biomolecules. These models, in
turn, allow us to elucidate the existence and functionality of
cellular machinery designed to alter the topology of circular DNA {\em
in vivo}. For example, due to the action of topoisomerase II enzyme
\cite{magnasco,roca}, it is known that the fraction of knotted plasmid
DNA in vivo (wild-type E. Coli) is much smaller than the statistical
equilibrium value predicted theoretically \cite{cozzarelli}. Besides
the occurrence of knots in DNA molecules that are free in solution
another biologically relevant case is the the presence of knots in
tightly-packed DNA, such as that confined in viral capsids
\cite{arsuaga,arsuaga2,harvey,harvey2,cristian,dewit2}. For example,
recent experimental investigations have revealed that the probability
of occurrence of cyclised knotted DNA is very high inside the T4
deletion mutant viral capsid \cite{arsuaga}. The experimental
characterization was accompanied by numerical simulations where a DNA
molecule was schematized as a phantom (i.e. interpenetrable) ring
which suggested that the observed high knotting probability was likely
to be a passive consequence of the spatial confinement, rather than to
be due to active biological mechanisms~\cite{arsuaga}.  In this work
we extend the previous studies~\cite{arsuaga,wiegel1,enzo2,Mansfield}
not only by considerably extending the length and degree of
confinement of the rings, but also by using the entire array of knot
invariants to identify the various types of occurring knots. This is
accomplished by first using a powerful statistical mechanical
framework for the efficient sampling of the ring conformations
confined in tight spheres and, secondly, by simplifying the sampled
ring conformations so as to aid their correct identification in spite
of their initially very complex two-dimensional projections.

The techniques used in this study to overcome these problems are
discussed in the next section. In sections 3 and 4 we present the
results pertaining to the knotting of a confined phantom chain and to
its writhe and chirality as it is confined in progressively smaller
spheres. We also discuss the scaling laws of some knotting
probabilities, and we map a diagram in which we identify the most
populated knot type for a given length and confinement level.  Section
5 contains a discussion of our results and our conclusions.

\section{Model and methods}

At the heart of this study is the generation, by stochastic
simulation, of uncorrelated configurations of random polymer rings of
various length and enclosed in spheres of various radii. As customary,
the ring is constituted by $N$ segments of unit length; no
self-avoidance is imposed on the ring. With reference to the origin of
the embedding in three-dimensional Euclidean space, each ring
conformation, $\Gamma$, is described by the coordinates of the
vertices, $\Gamma \equiv \{ \vec{r}_1, \vec{r_2}, \vec{r}_N,
\vec{r}_{N+1} \equiv \vec{r}_1\}$. To each ring configuration we
associate a radius $R(\Gamma)$, defined as $R = \max_i |\vec{r}_i|$
which measures the distance from the origin of the vertex that is
farthest from the origin itself. Obviously, a ring $\Gamma$ can be
enclosed only in spheres having radius larger than $R(\Gamma)$. This
ensemble is therefore different from the one where the centre of the
enclosing sphere is chosen to coincide with the geometrical centre of
the ring. Although this second alternative is also intuitive and
natural, the ensemble chosen here seems most naturally related to the
experimental situation where one has the hull (e.g. a viral capsid)
fixed in space and the enclosed molecule occupying any internal
region.

We aim at calculating, for any given ring length, $N$, the number
density of ring conformations of given knot type, $K$, that are
generated by random sampling from the ensemble and can be accomodated
inside a sphere of radius $R$. This quantity will be indicated as
$W_N(K,R)$. A technique based on multiple Markov chains and histogram
reweighting will be used to determine $W_N(K,R)$ up to a
multiplicative constant \cite{cristian,lesdiablerets,hirudin}. The
value of this constant is irrelevant for calculating the occurrence
probability of various knots types (which is expressed in terms of
suitable ratios of $W_N$ terms, as discussed later).

In principle $W_N(K,R)$ could be obtained by an unbiased sampling of
random ring conformations. In practice, the overwhelming majority of
polymer rings will be ``swollen''\cite{Millet} and, therefore, an impractically
long computing time would be necessary to accumulate reliable
statistics for highly confined conformations. A more efficient
alternative is to generate a succession of conformations picked with
importance sampling criteria. At any stage of the procedure, one
deforms the current ring conformation stochastically by moves
preserving the chain connectivity and bond length. The usual
Metropolis scheme is then employed to reject or accept newly-generated
conformations based on the score assigned to them. In the simplest
approach, the scoring function is chosen so as to penalize severely
cases where a preassigned threshold value, $\bar{R}$ is exceeded
(while no penalty is introduced for configurations satisfying the hull
constraint). Starting from an arbitrary conformation the stochastic
evolution will eventually drive the system to configurations that can
fit inside the preassigned enclosing hull. However, this method also
presents difficulties since, for the entropic reasons mentioned above,
most of the sampled configurations will attain the maximum allowed
value of $R$, that is $\bar{R}$. Their stochastic modification is
therefore likely to produce a value for $R$ that violates the hull
constraint and hence leads to rejection. The large numbers of
rejections encountered is such that impractically large number of MC
steps are required to decorrelate the system as $\bar{R}$ is
decreased.

In the present study we reduce the impact of these sampling
difficulties by two means. Rather than working in the ensemble of
fixed $\bar{R}$, we work in the conjugated ensemble by introducing an
auxiliary parameter, $P$, akin to the familiar ``hydrostatic''
pressure. For a given value of $P$ and $N$, the number of 
{\em sampled} polymer
rings having knot type $K$ and radius equal to $R$ is proportional to
the weight $\exp\left(-P\,R\right)$ and to the number 
density of
conformations having knot type K and radius equal to $R$, $\tilde{W}_N(K,R)$
(the logarithm of $\tilde{W}_N(K,R)$ provides the configurational
entropy of the rings up to an additive constant):

\begin{equation}
n_N(K,R) \propto \tilde{W}_N(K,{R})\, e^{-P\,R}
\end{equation}

The tilde superscript is used to distinguish this quantity from the
one introduced before which denoted the number density of conformations of
given knot type that have radius less than or equal to $R$ (as opposed to
having exactly radius $R$). Since

\begin{equation}
W_N(K,\bar{R}) \propto \int_0^{\bar{R}}\, dR \, \tilde{W}_N(K,R)
\end{equation}

\noindent we have

\begin{equation}
W_N(K,\bar{R}) \propto \int_0^{\bar{R}}\, dR\, n_N(K,R)\,  e^{+P\,R}
\end{equation}

\noindent At each value of $P$ one can therefore reconstruct
$W_N(K,R)$ throughout the range of explored values of $R$. As $P$ is
increased lower values of $R$ are encountered. By optimally
superimposing the $W_N(K,R)$ obtained at different values of $P$ one
can then recover the value of $W_N(K,R)$, for a continuous
range of $R$ spanning from the lowest to the largest observed values of
$R$. The optimal superposition of the various $W$'s is carried out
using the standard Ferrenberg-Swendsen approach. Once $W_N(K,R)$ is
known, various quantities of interest can be calculated. For example,
at given length, $N$, the expected
fraction of polymer rings of knot type $K$
that can fit inside a spheres of radius $R$ is given by:

\begin{equation}
P_N(K,R) = {W_N(K,R) \over \sum_{K^\prime} W_N(K^\prime,R)  }
\label{eqn:prob}
\end{equation}

\noindent where $K^\prime$ spans all knot types.

\noindent The reweighting procedure was used to produce most of the
data presented here. The error in the quantities (\ref{eqn:prob}) and
(\ref{eqn:prob2}) is estimated from the semidispersion observed by
applying the weighted histogram method separately to the first and
second half of the polymer rings generated stochastically. Moreover
the data pertaining to various ``pressures'' were not collected from
independent runs, but by using the standard, yet powerful,
multiple-Markov chains scheme.  Within the latter framework all
``replicas'' of the system at the various pressures are run in
parallel and, at times greater than the largest autocorrelation time,
one proposes swaps of the ring conformations among two replicas at
nearby pressure values. The swap is accepted according to a
suitably-generalized Metropolis criterion \cite{enzo1}. The exchange
of the replicas results in a significant decrease of the
autocorrelation time in the system, and hence in a more effective
sampling of the accessible conformational phase space.

The final focus of this section is the description of the strategy
used to identify the type of knot associated to any given ring
conformation. The classification of the knot type of a given ring
conformation is facilitated by reducing the number of crossings in the
knot diagram, while preserving knot type.  Randomly generated rings,
especially those confined in small spheres, typically present
projections with a number of crossings that vastly exceeds the minimal
one for that knot type. For example, unknotted conformations in rings
of $N=400$ segments confined in spheres of radius $R=10$ typically
present $\sim 400$ crossings. No general deterministic algorithm at
present exists for obtaining minimal projections starting from a
generic suboptimal one. A few stochastic methods have, however, been
introduced that can simplify the initial diagram considerably.  Our
strategy to simplify the knot structure was based on a generalization
of the techniques of refs.~\cite{muthukumar}
and~\cite{taylor}. Starting from an initial conformation, we pick at
random one vertex $i$ and perform with equal probability one of the
following two moves:

\noindent The first move consists of assigning a new position for $i$
obtained as $\vec{x}^{new}_i = \alpha\, \vec{x}_i + (1 - \alpha)
(\vec{x}_{i-1} + \vec{x}_{i+1})/2$, where $\alpha$ is a random number
picked uniformly in the interval [0.2,0.8]. The proposed new position
is accepted only if the deformation of the chain when vertex $i$ is
moved continuously from the old to the new position does not lead to
self-intersections of the ring. The repeated application of the
operation leads to a smoothing and contraction of the initial chain
and hence a simplification of the crossing pattern~\cite{taylor}.

\noindent The second type of operation is an attempt to simplify the
chain by removing vertex $i$. This can be viewed as a special case of
the previous move where $\alpha$ is set equal to zero, i.e. vertex $i$
is made collinear with the previous and following vertices. If this
move is acceptable (in the same sense as before) then $i$ is removed
from the list of vertices.

\begin{figure}
\begin{center}
\vskip 4.truecm
\includegraphics[width=3.5in]{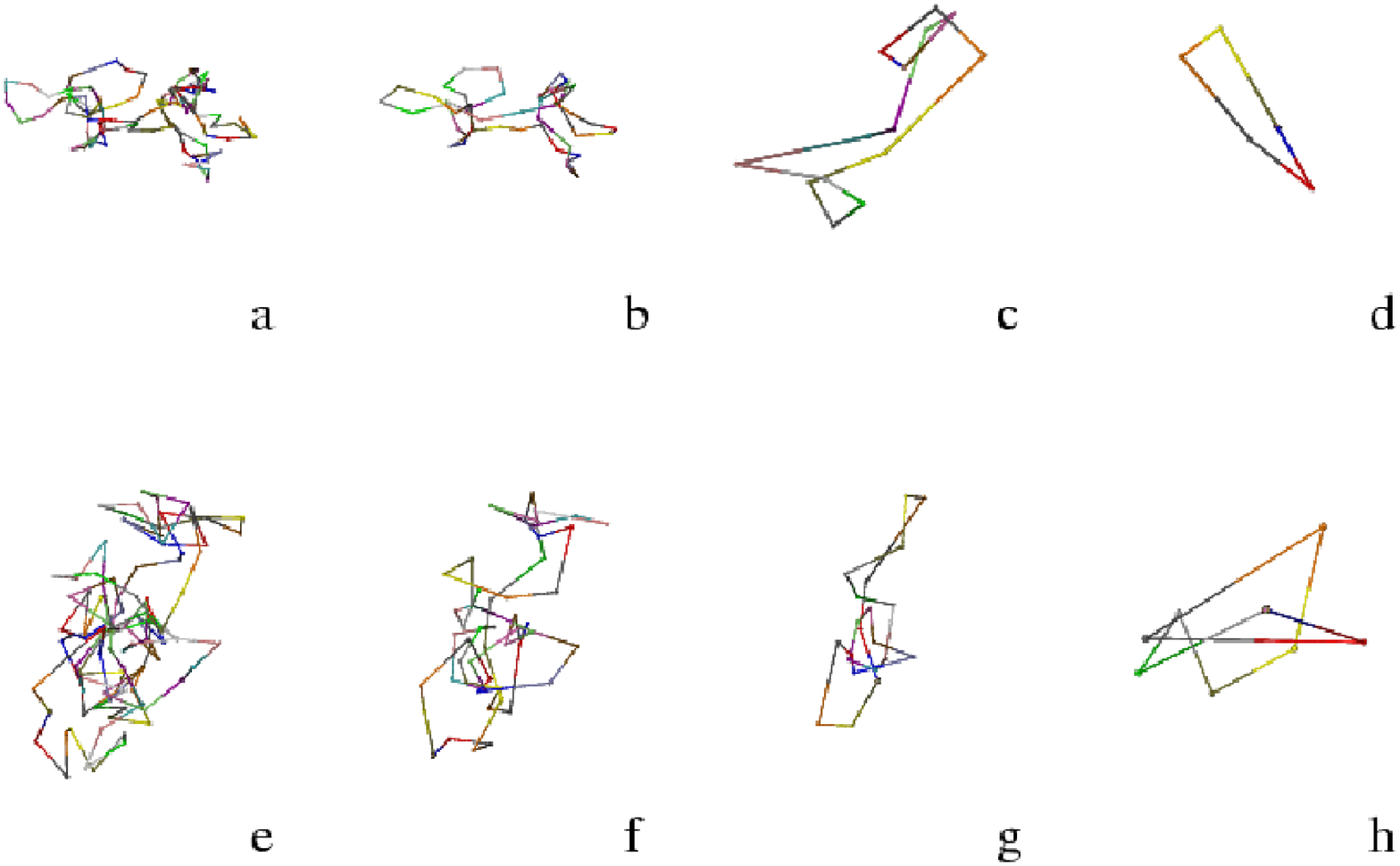}
\end{center}
\caption{}
\label{smoothing}
\end{figure}

The smoothing and shrinking operations are attempted until the number
of vertices does not decrease in 10 successive system sweeps (a sweep
consists in proposing either of the two elementary moves for a number
of randomly chosen vertices equal to the ring length). Convergence is
typically achieved in a dozen sweeps. This simplification procedure
can reduce dramatically the number of crossings encountered in an
arbitrary projection as shown in Fig.~\ref{smoothing}. For example,
rings with $N=400$ segments inside spheres of radius $R=10$ are
typically simplified down to rings with $N^\prime \sim 35$
(non-equilateral) segments with  projections having, on
average, $30$ crossings.

Of the simplified ring configurations we chose the projection with the
minimum number of crossings produced by the smoothing scheme and
encoded it in terms of the Dowker code~\cite{dowker} and then fed this
code to the Knotfind programme~\cite{knotfind}.  The Knotfind
programme then attempts to further reduce the number of crossings by
performing modifications on the code (the modifications include some
Reidemeister moves).  The resulting knot representation is compared to
a library of prime knot representations for correct identification. By
these means it was possible to precisely identify prime knots (or
prime components of composite knots) for prime and composite knots of
up to 16 crossings. In some cases, even after the hierarchy of
reduction of knot complexity, the diagrams could not be
identified. This could reflect the genuine fact that the knot under
consideration had a minimal projection with more than the threshold
number of crossings (i.e. 16), or could simply reflect the inability
to simplify it to a point where the knot projection had 16 or fewer
crossings.

\begin{figure}
\begin{center}
\vskip 4.truecm
\includegraphics[width=4.0in]{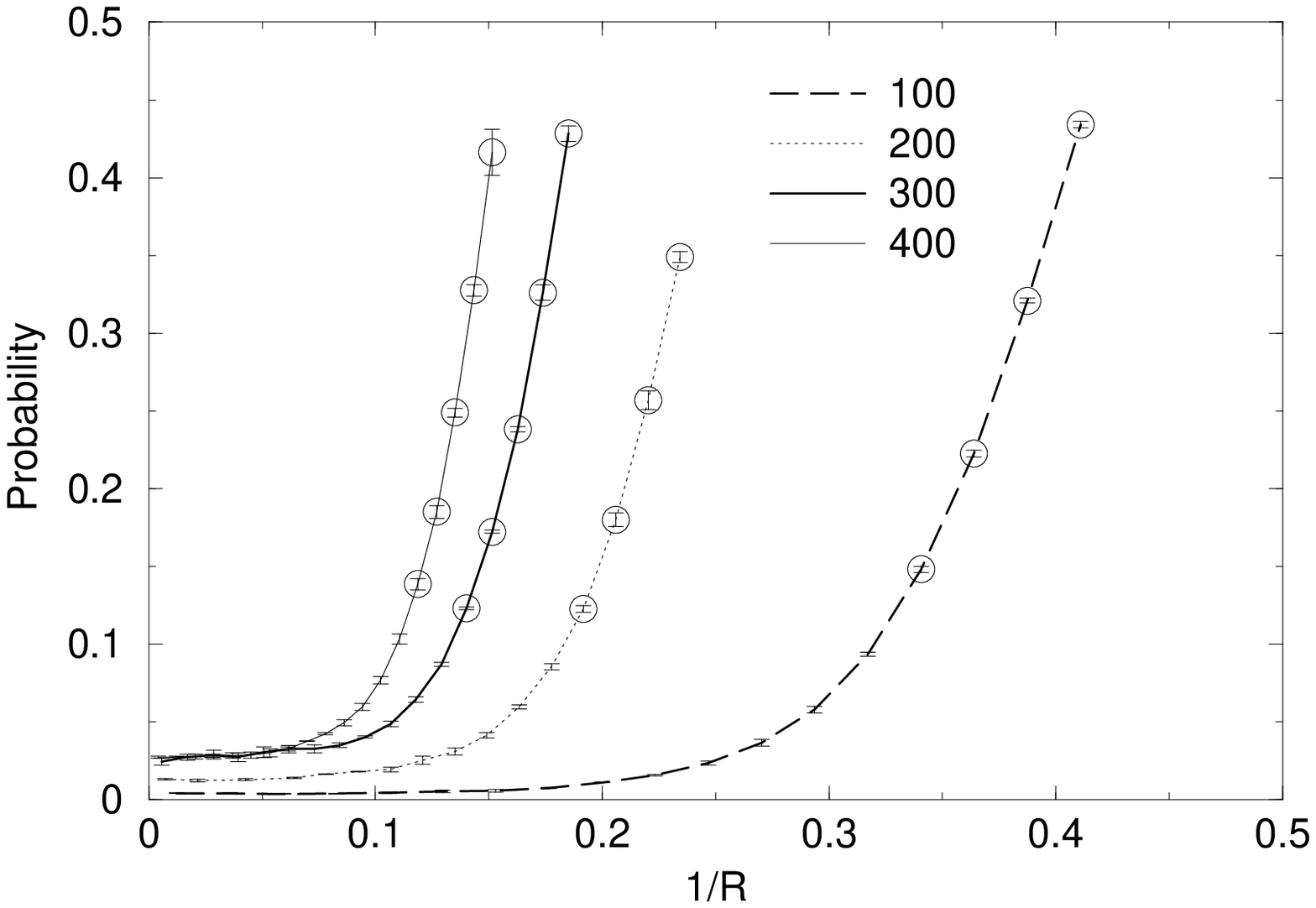}
\end{center}
\caption{}
\label{fig:unknownknots}
\end{figure}

Whenever this situation is encountered the knot is classified as
``unknown''. As visible in Fig. \ref{fig:unknownknots} the number of
``unknown'' knots grows with the decrease of the confining hull, $R$,
as expected intuitively.

\begin{figure}
\vskip 5.truecm
\begin{center}
\includegraphics[width=3.5in]{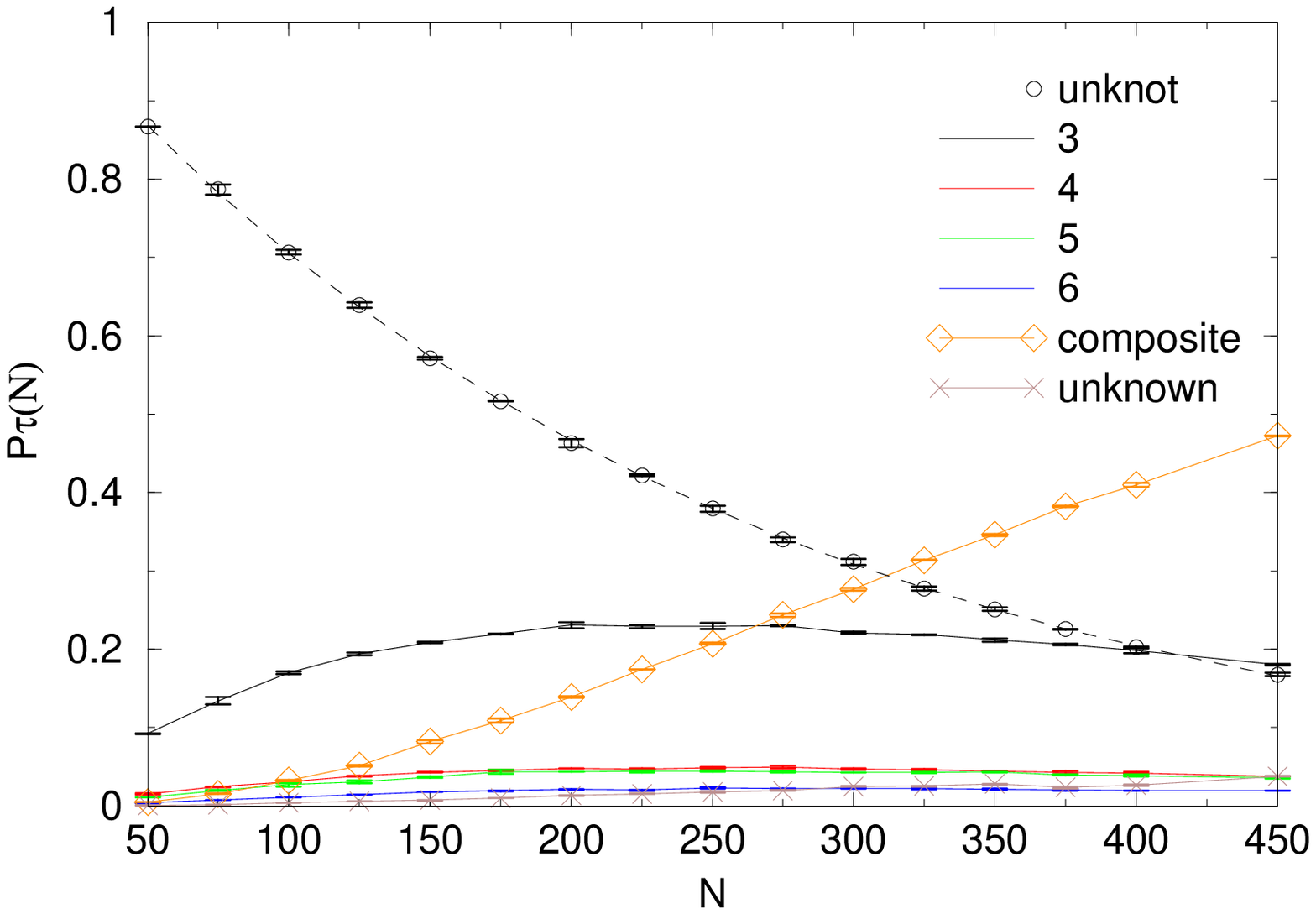}
\end{center}
\caption{}
\label{infR}
\end{figure}

For the uncontrained case, $R=\infty$, the occurrence probability,
$P_{\tau}(N)$ of various types of knots, $\tau$, is shown in
Fig.~\ref{infR}. The results appear consistent with previous studies
~\cite{muthukumar,wiegel1,stasiak,stasiak2,deguchi} and the well known
exponential decay~\cite{stu,pippenger,diao,diao2} of $P_{\tau}(N)$ as
a function of $N$ is verified for the simplest prime knots. In
particular, by fitting of the unknot curve with the a
single-exponential decay $P_{\tau} \propto \exp{-(N/N_c)}$ we obtain
the decay length $N_c = 244 \pm 5$ which agrees quite well with the
up-to-date independent estimates~\cite{rawdon05,deguchi}. 

\section{Results for confined rings}
\begin{figure}
(a)\includegraphics[width=3.0in]{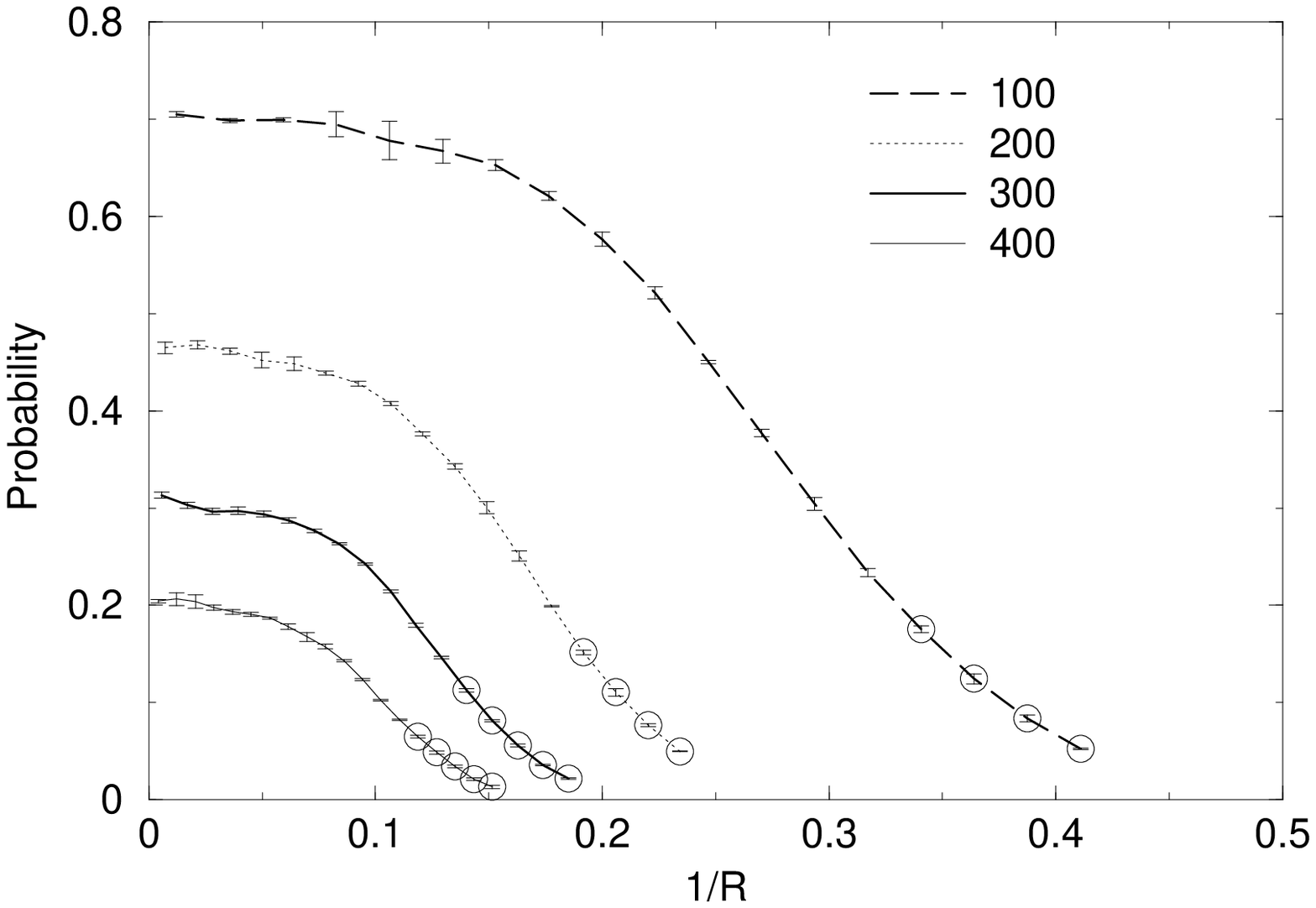} (b)\includegraphics[width=3.0in]{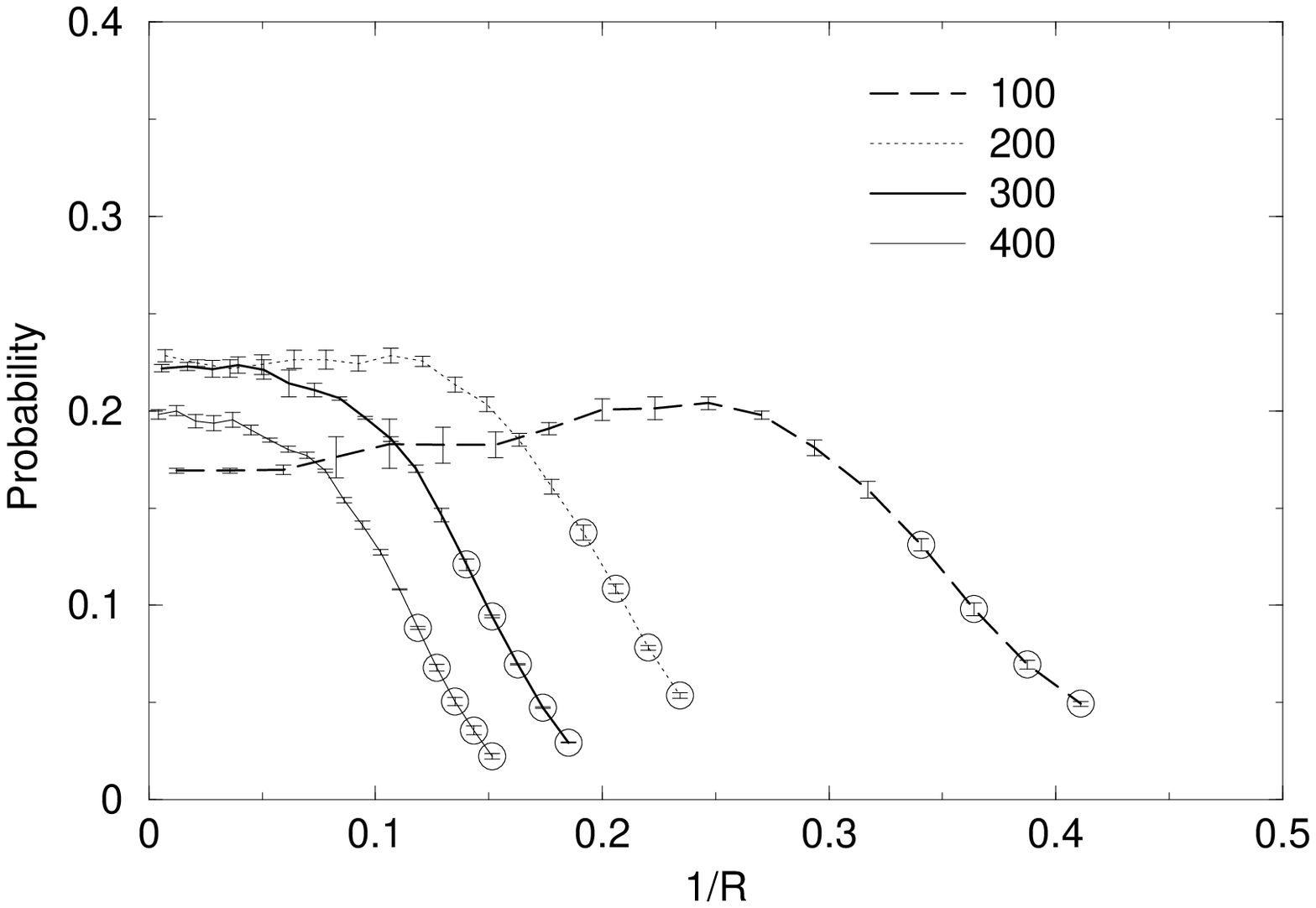}\\
(c)\includegraphics[width=3.0in]{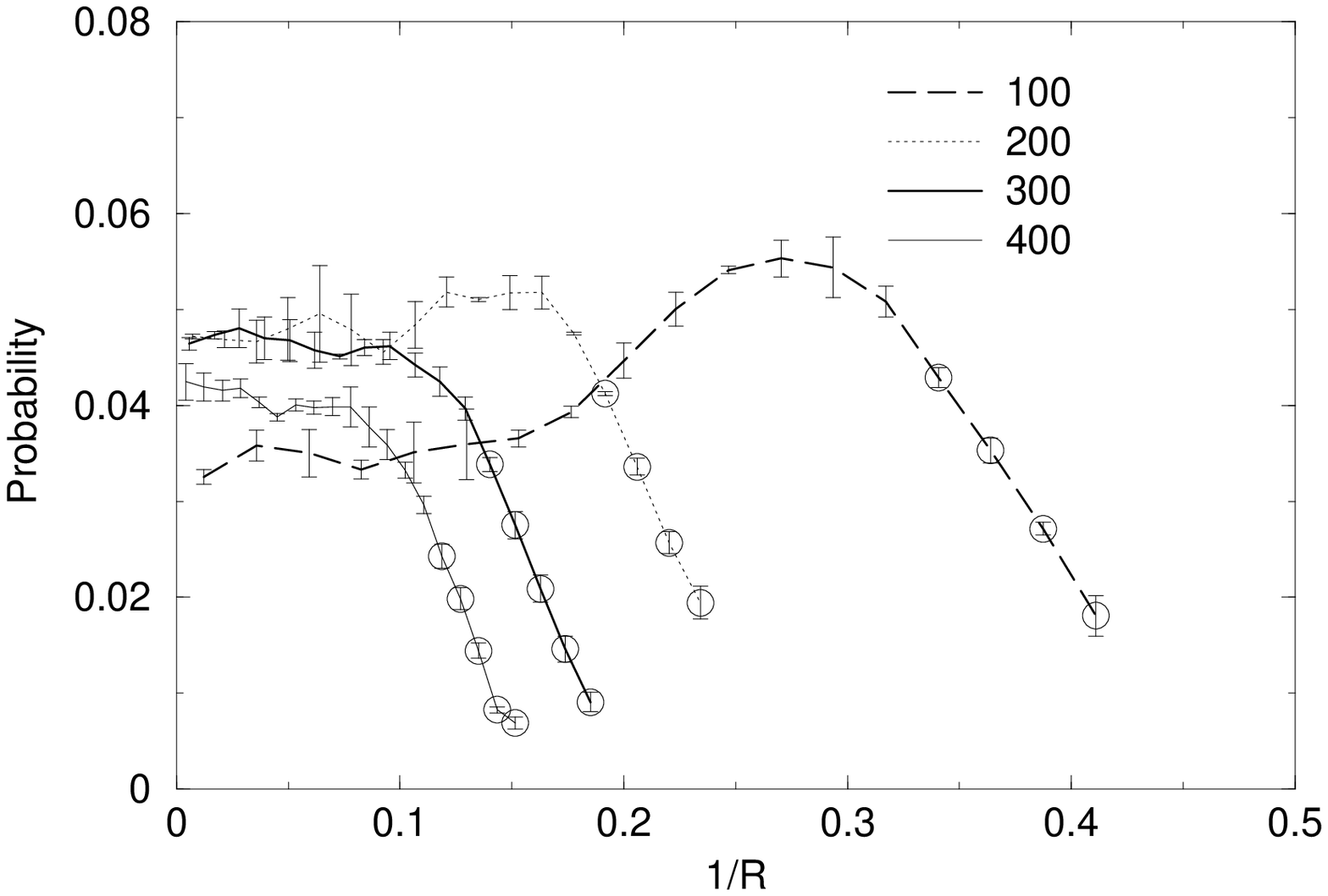} (d)\includegraphics[width=3.0in]{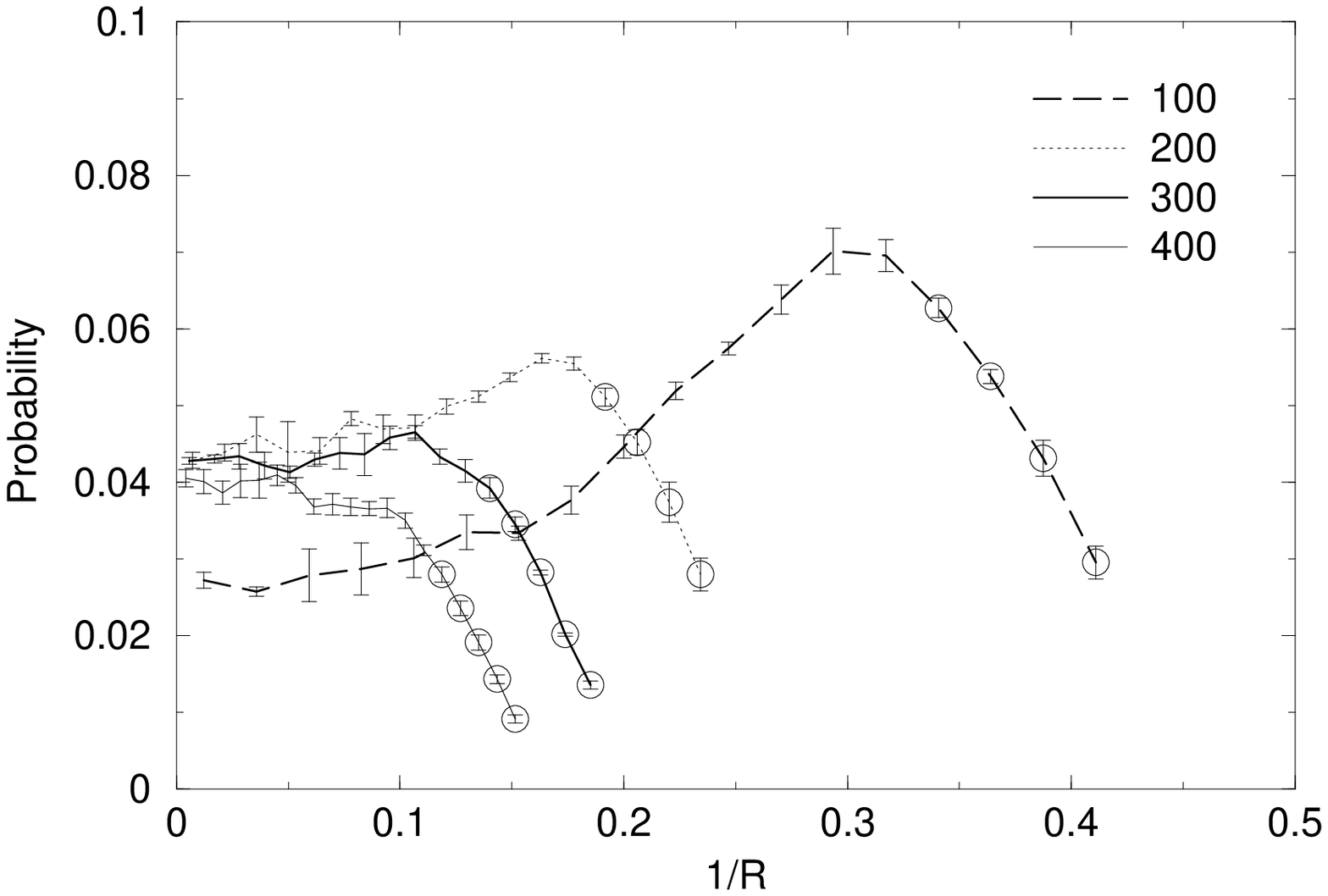}\\
(e)\includegraphics[width=3.0in]{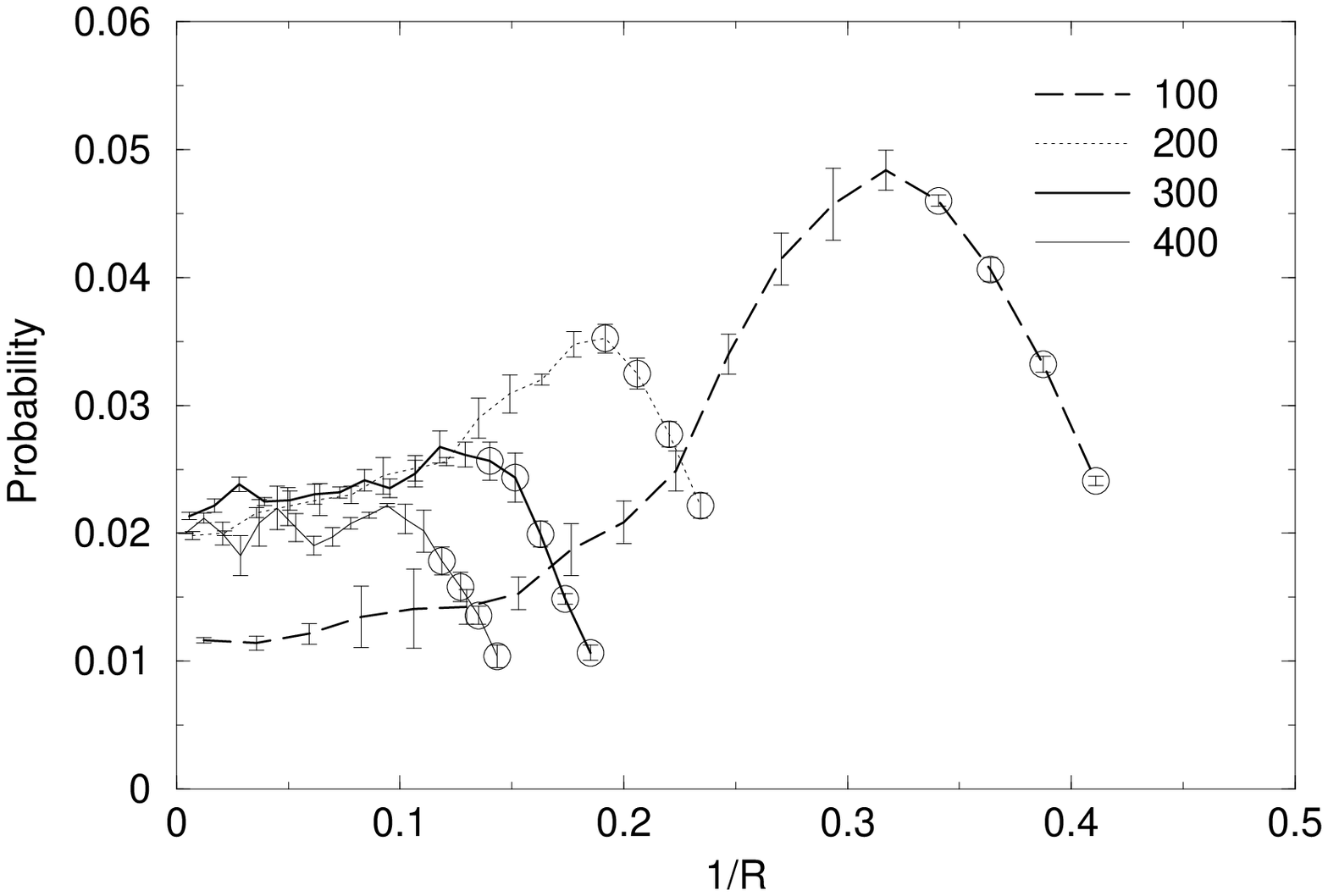} (f)\includegraphics[width=3.0in]{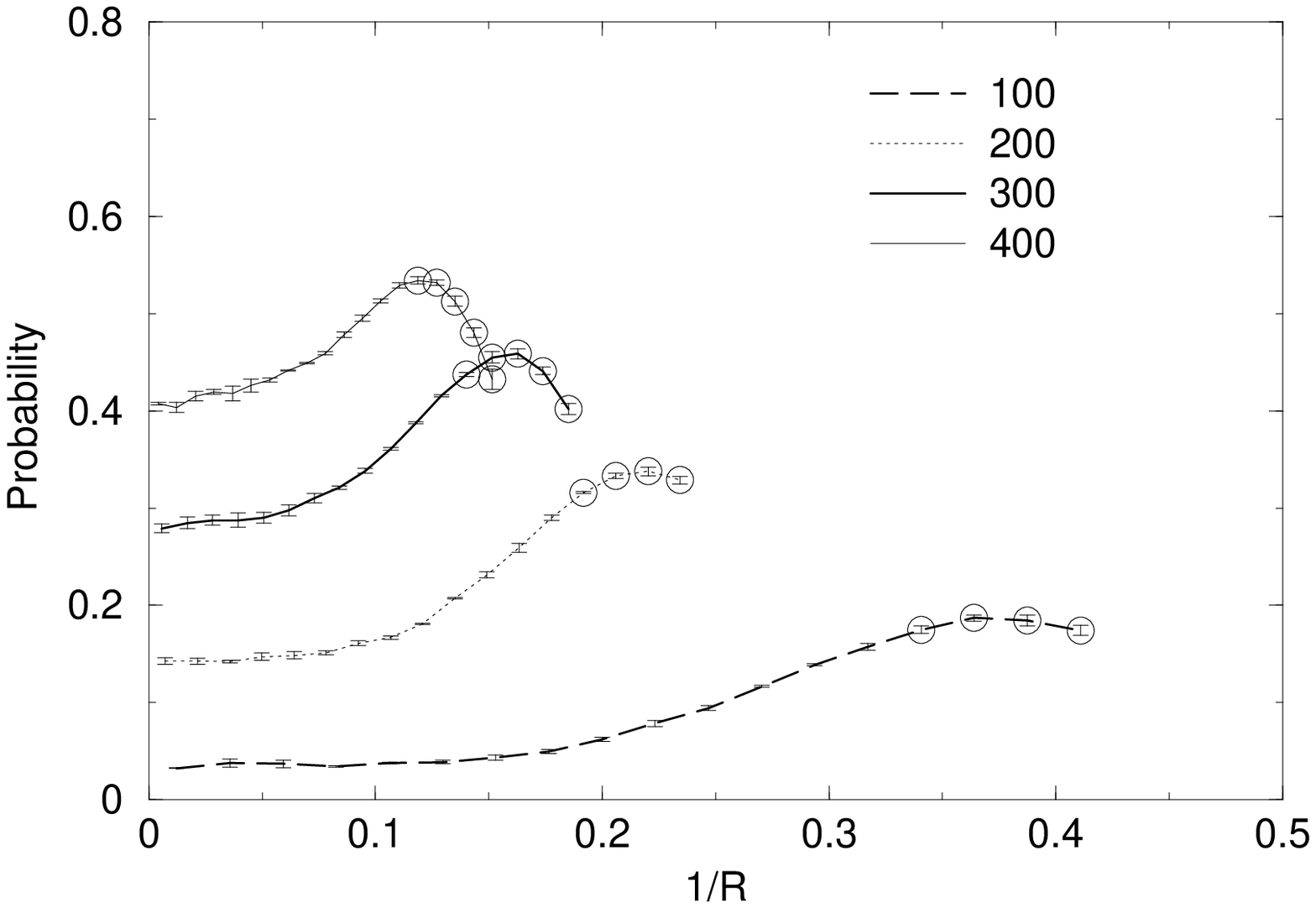}
\caption{}
\label{knotprob2}
\end{figure}

\begin{figure}
(a)\includegraphics[width=3.0in]{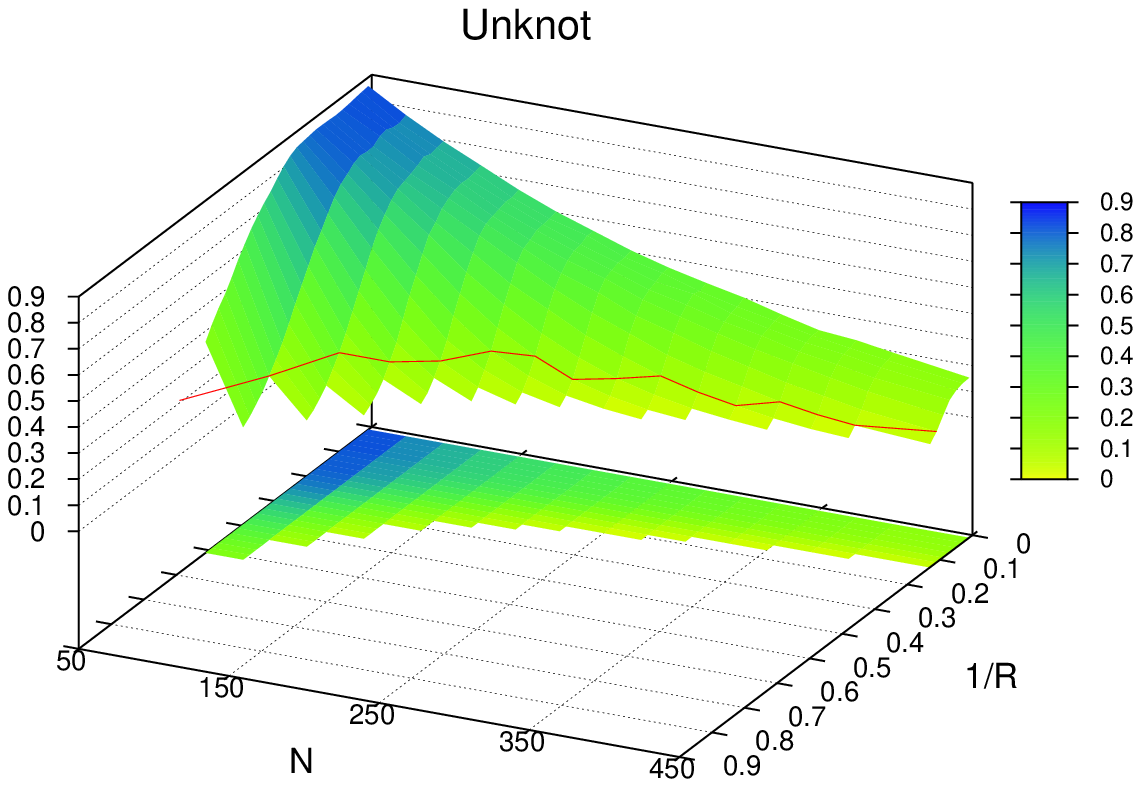} (b)\includegraphics[width=3.0in]{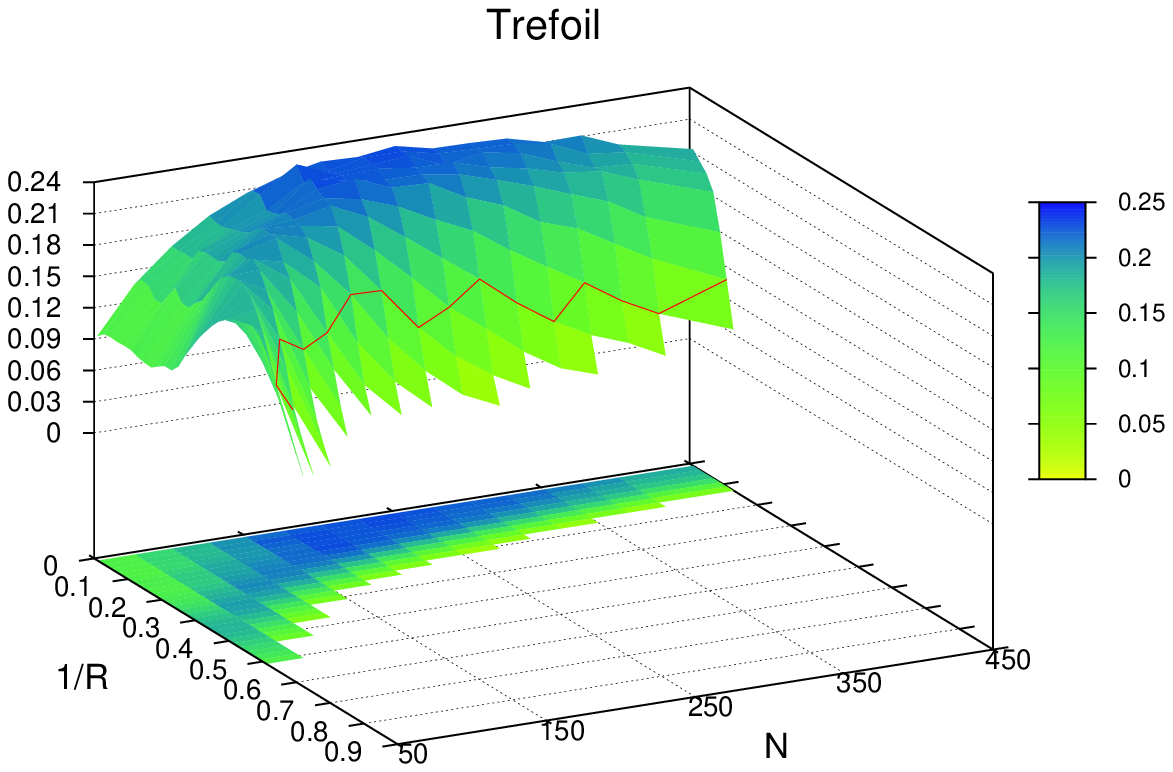}\\
(c)\includegraphics[width=3.0in]{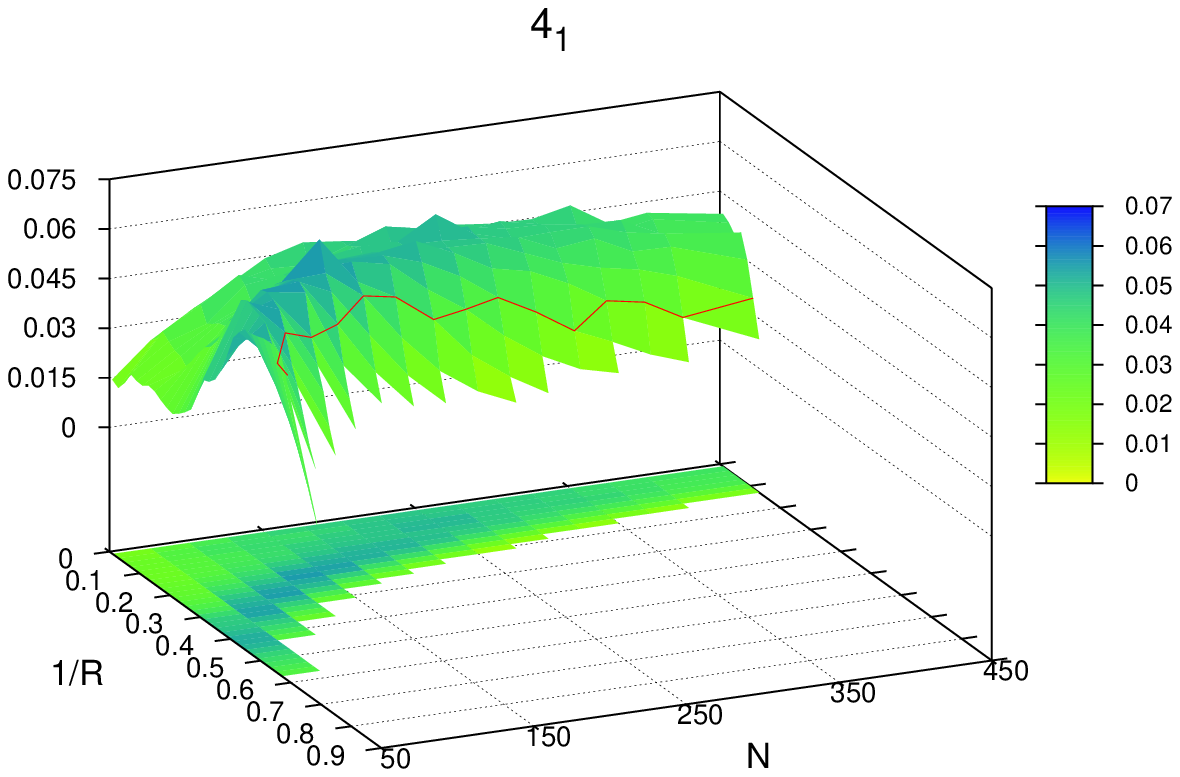} (d)\includegraphics[width=3.0in]{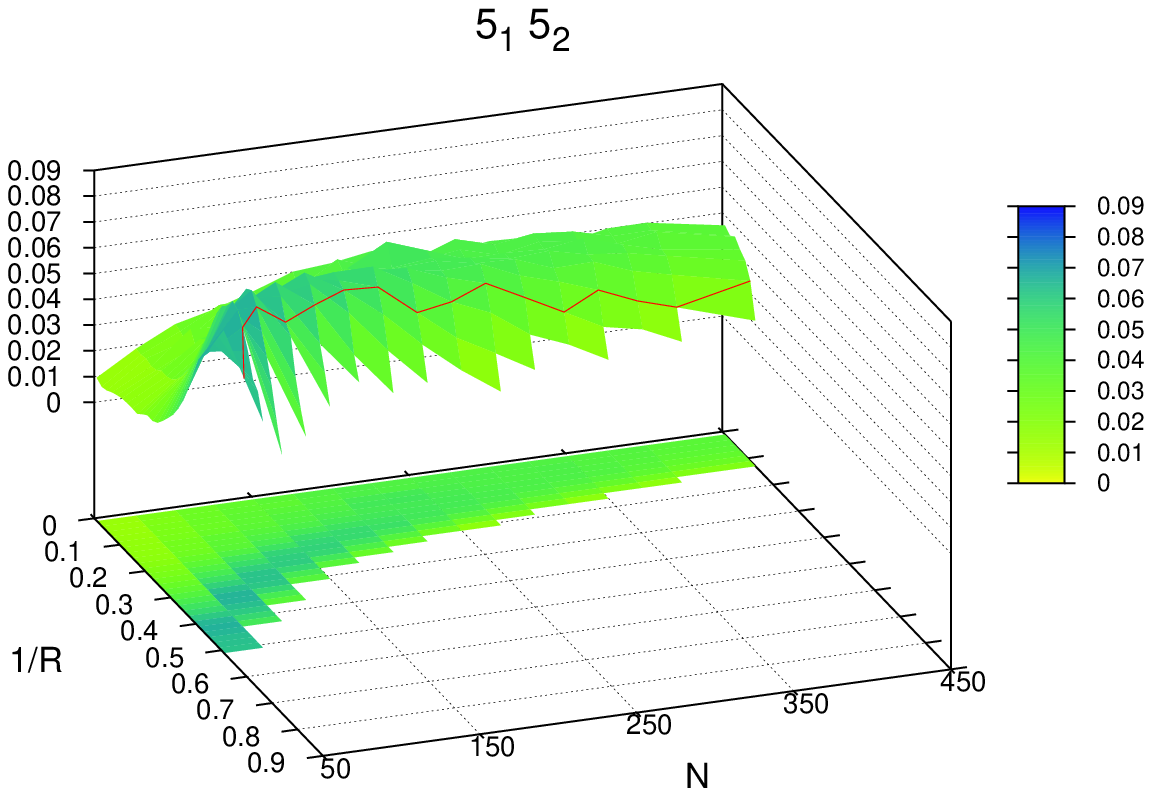}\\
(e)\includegraphics[width=3.0in]{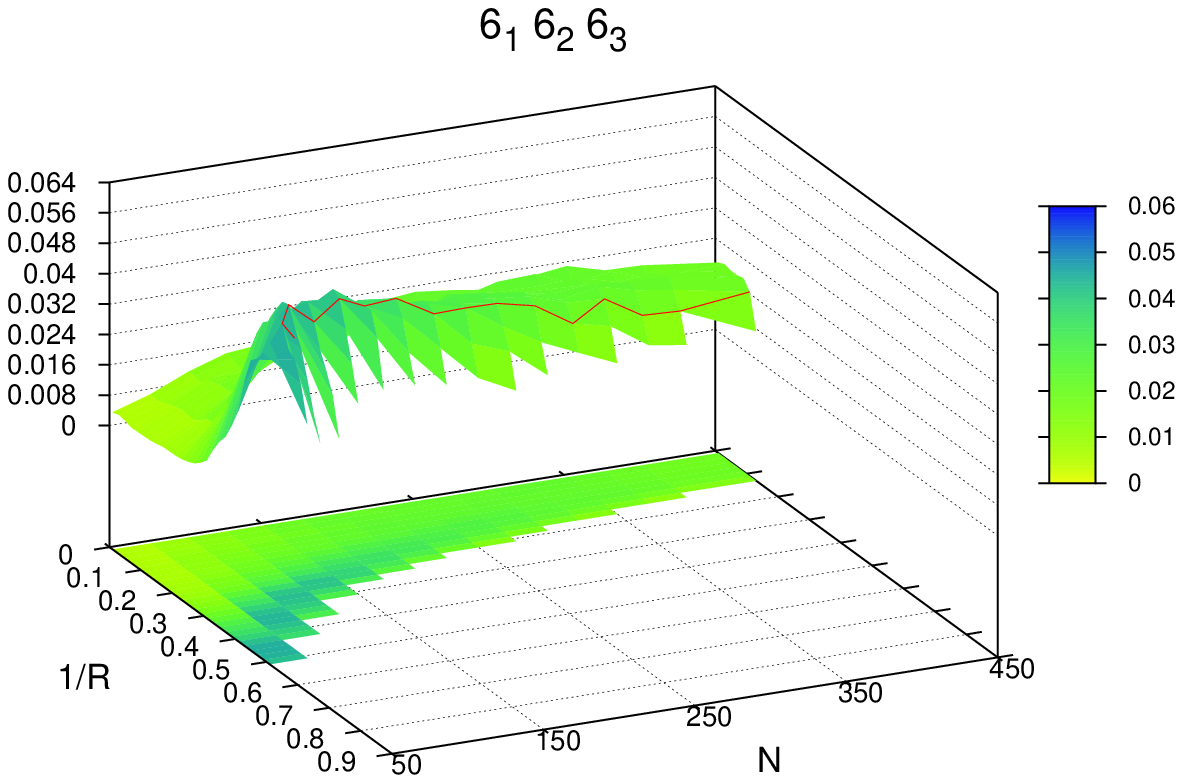} (f)\includegraphics[width=3.0in]{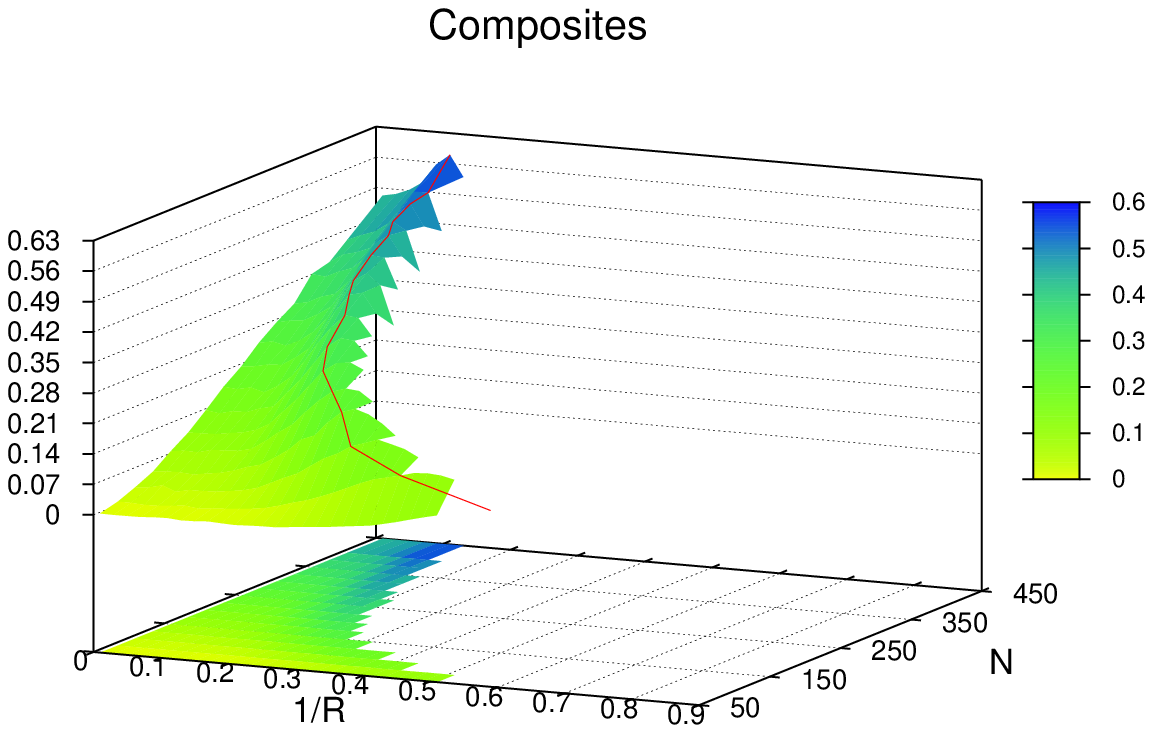}
\caption{}
\label{knotprobs}
\end{figure}

We now discuss the case of random rings confined within a sphere of
radius $R$. In Figs.~\ref{knotprob2}a and \ref{knotprobs}a we show the
unknotting probability as a function of inverse radius for rings of
$N$ segments. Besides the statistical error reflecting the finite
sampling of knot configurations, the confidence in the curves for the
occurrence probability is affected by the fraction of ``unknown''
knots, which may contain 
knots with crossing numbers less than or equal to 16, as well as knots with 
crossing numbers which exceed 16.
As one moves towards smaller confining radii,
the fraction of unknown knots becomes progressively higher, thus making
uncertain the precise quantitative estimate of the occurrence of the
various knot types. We use a threshold of 10\% for the occurrence of
unknown knots, $P_{unknown}$, in order to separate the reliable from
the unreliable parts of the probability profiles. Moreover, no data are
presented when $P_{unknown}$ exceeds 50\%. To distinguish graphically
these cases in Figs.~\ref{fig:unknownknots}-~\ref{knotprob2} the points 
falling in the region
where $P_{unknown} > 0.1$ are denoted with open symbols. The
reliable-unreliable boundary is instead shown as a red line on the
two-dimensional probability surface of Fig.~\ref{knotprobs}.

\begin{figure}
\vskip 5.truecm
\begin{center}
\includegraphics[width=4.0in]{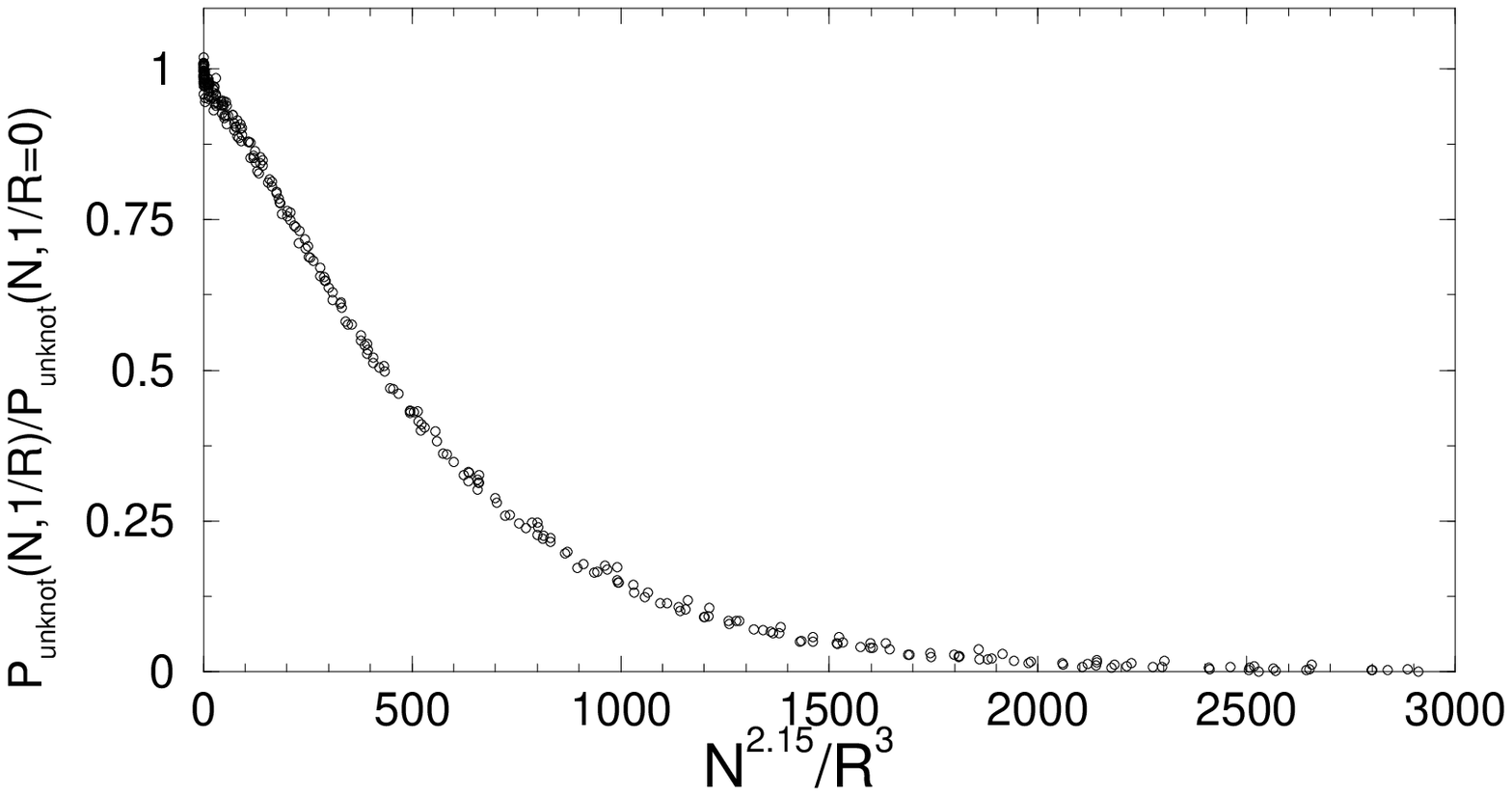}
\end{center}
\caption{}
\label{scal_unknot}
\end{figure}

Different curves correspond to different values of $N$ ranging from
$50$ up to $450$ with incremental step $\Delta N= 25$. Note that, for
fixed $N$, the unknotting probability remains fairly constant for
$R>R_c$ ($R_c$ is a length dependent threshold) and
undergoes a pronounced decrease for $R<R_c$. 
For fixed $R$, the knotting probability decreases with $N$ as expected. 
Michels and
Wiegel, in their pioneering work \cite{wiegel1}, analysed the scaling
properties of the unknotting probability for moderate values of $N$
and confining radii, $R$, and suggested the following scaling form
\begin{equation}
{P_{unknot}(N,{1 \over R})
\over P_{unknot}(N,{1\over R}=0)} =  g({N^\alpha \over
  R^3})\ ,
\label{eqn:scaling}
\end{equation}

\noindent where their estimate for the exponent $\alpha$ was 2.28. 
We have analysed our data, which span over much
larger values of $N$ and confining degree than ref. \cite{wiegel1},
and established that, though the scaling law of eqn. \ref{eqn:scaling}
remains valid, the exponent $\alpha$ is noticeably different from the
previous estimate. The value of $\alpha$ providing the best collapse
of the unknotting profiles for $N \ge 100$ was obtained from the
procedure of refs. \cite{flavio_somen,zillio} and resulted equal to
$\alpha = 2.15 \pm 0.04$. The uncertainty on the exponent was
estimated by dispersion of the optimal values of $\alpha$ leading to
the curves collapse for two distinct sets of values of $N$. The
quality of the obtained collapse is visible in Fig.~\ref{scal_unknot}.

\begin{figure}
\phantom{.}
\vskip 5.truecm
\begin{center}
\includegraphics[width=4.0in]{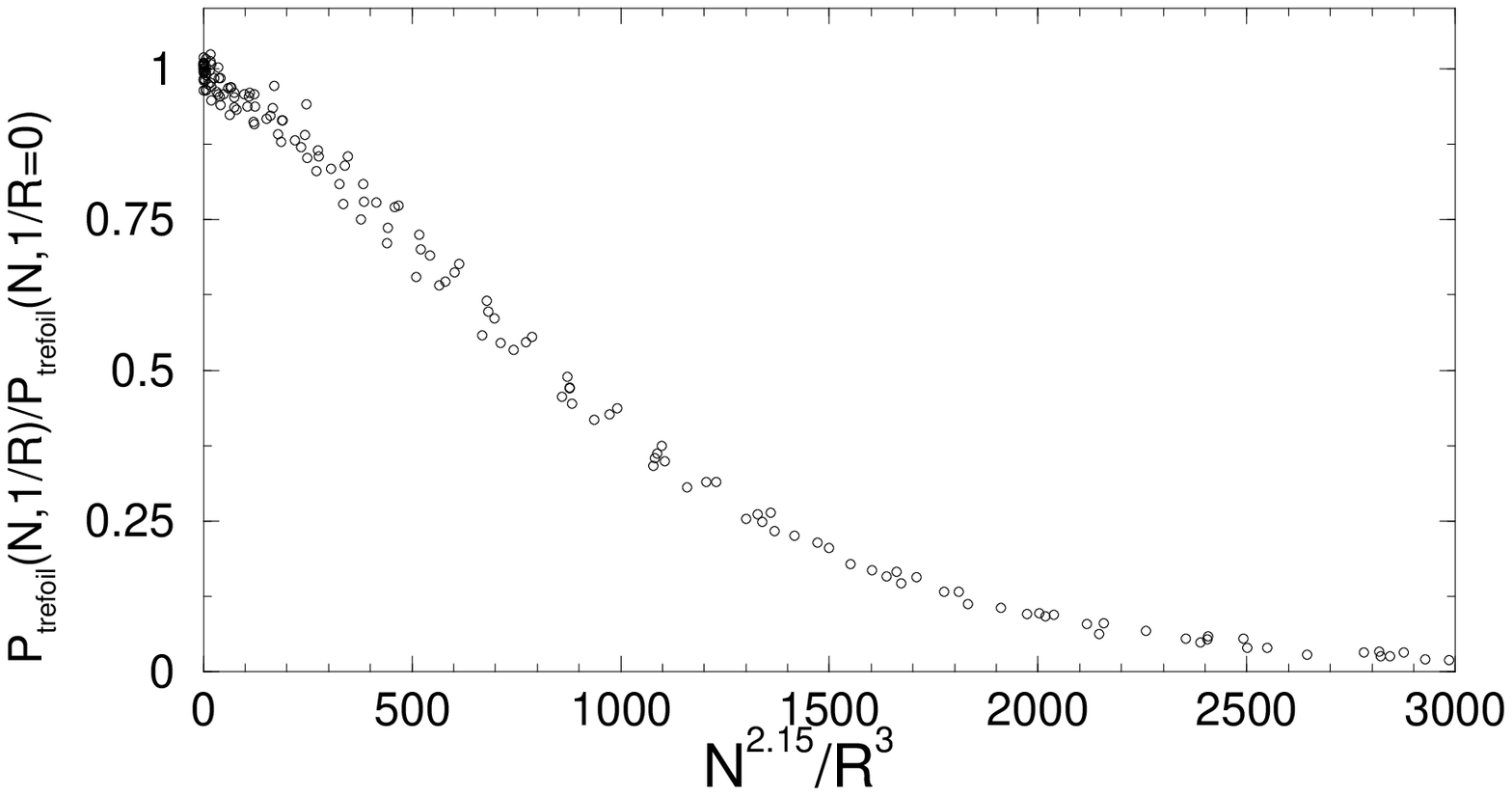}
\end{center}
\caption{}
\label{scal_tref}
\end{figure}

Figs.~\ref{knotprob2}b and \ref{knotprobs}b show, instead, the $R$
dependence of the probability to observe a trefoil knot ($P_{3_1}$)
for various lengths of the polymer ring. As for the unknotting
probability there is a range of $R$ ($R>R_c$) for which $P_{3_1}$ does
not change too much with $R$. This ``plateau'' is more visible for
small values of $N$ when, for sufficiently small confining radii,
$P_{3_1}$ is a non-monotonic curve with one maximum. As the
confinement radius $R$ is further reduced, $P_{3_1}$ decreases in
favour of more compact conformations.  For longer polymers ($N>125$)
the maximum becomes progressively less evident and we observe a
shoulder for small values of $1/R$, that disappears for $N ~ 400$.
The possibility that the probability of trefoil knots also obeys a
scaling law has not (to our knowledge) 
been investigated before. Fig. \ref{scal_tref}
displays the data of trefoil probabilities plotted as a function of
the same scaling variable obtained for the unknot. The data, which
pertain only to lengths $ 250 \le N \le 450$ (since shorter rings
display the above mentioned ``peak'' or shoulder) appear well
collapsed. The optimal collapse of the trefoil data is obtained for an
exponent $\alpha \simeq 2.3$ somewhat larger than for the unknot (but the
trefoil data pertains to fewer ring lengths than for the
unknot). The results therefore indicate that, for sufficiently large
$N$, the trefoil probabilities also obey a scaling law with an
exponent that may be the same as that of the unknots.

\begin{figure}
\begin{center}
\vskip 5.truecm
\includegraphics[width=4.0in]{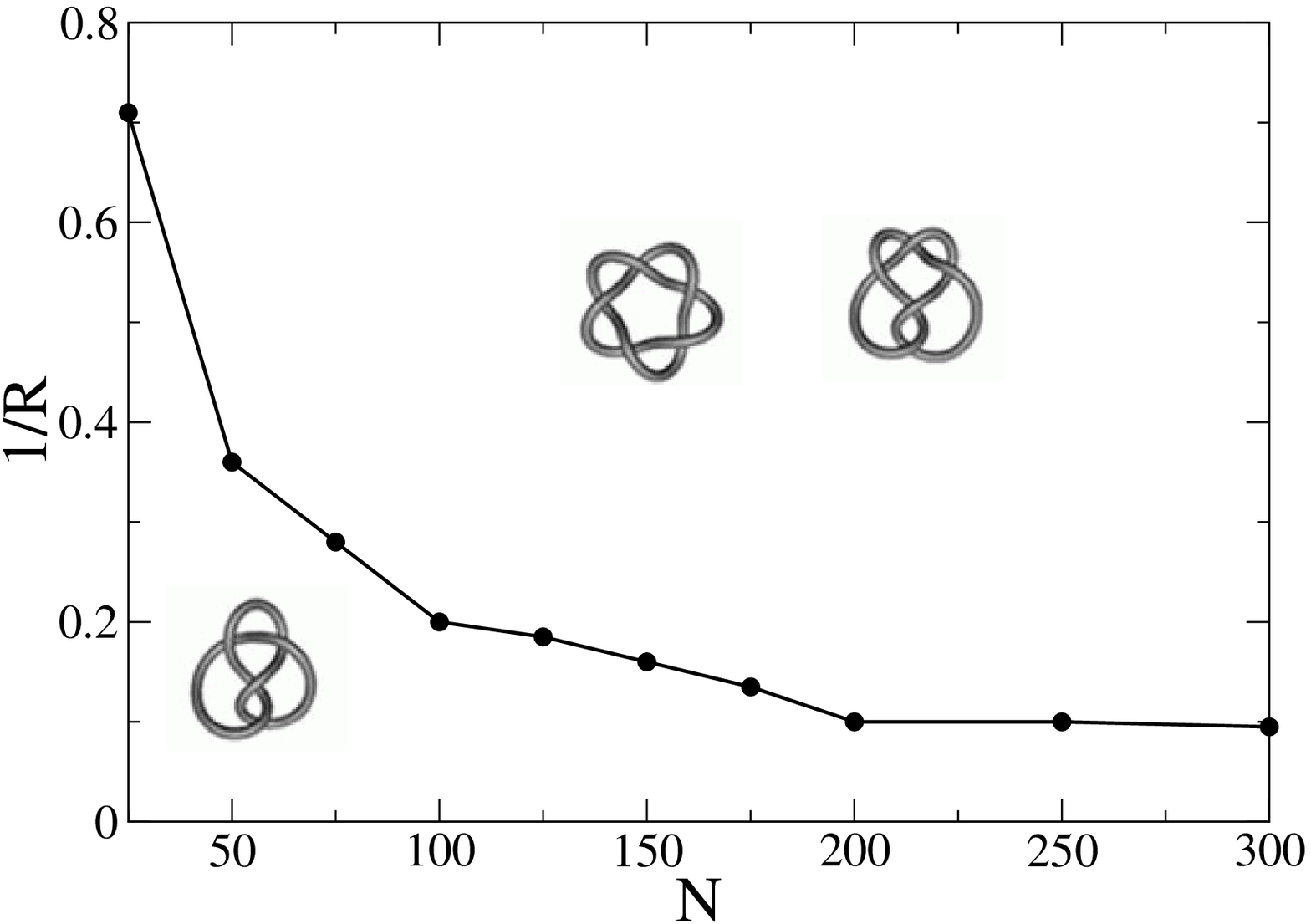}
\end{center}
\caption{}
\label{4v5}
\end{figure}

\noindent Figs. \ref{knotprob2}c,d,e and \ref{knotprobs}c,d,e and show
plots analogous to the one in figs.\ref{knotprob2}b and
\ref{knotprobs}b but now refer to the probability of forming $4$,
$5$ and $6$ crossing knots respectively. The trend observed for the
trefoil knot is also observed for the other prime knots considered
here. Indeed, all the curves for short polymers (small $N$)
have a maximum as a function of the inverse radius, while those
for longer polymers decrease monotonically with $1/R$. It is
interesting to observe that for polymers with $N$ up to $200$, the
confinement eventually forces the $5$ crossing knots to outnumber
the $4_1$ knots, which are more numerous in the unconstrained case
($R=\infty$).  This result is consistent with what observed in
Ref.~\cite{dewit2} for shorter random rings. This effect is
illustrated in Fig.~\ref{4v5} which clarifies how the confining radius
corresponding to the crossover from $4$ to $5$ crossing knots increases for
increasing length $N$.

\begin{figure}
\begin{center}
\vskip 2.truecm
\includegraphics[width=4.0in]{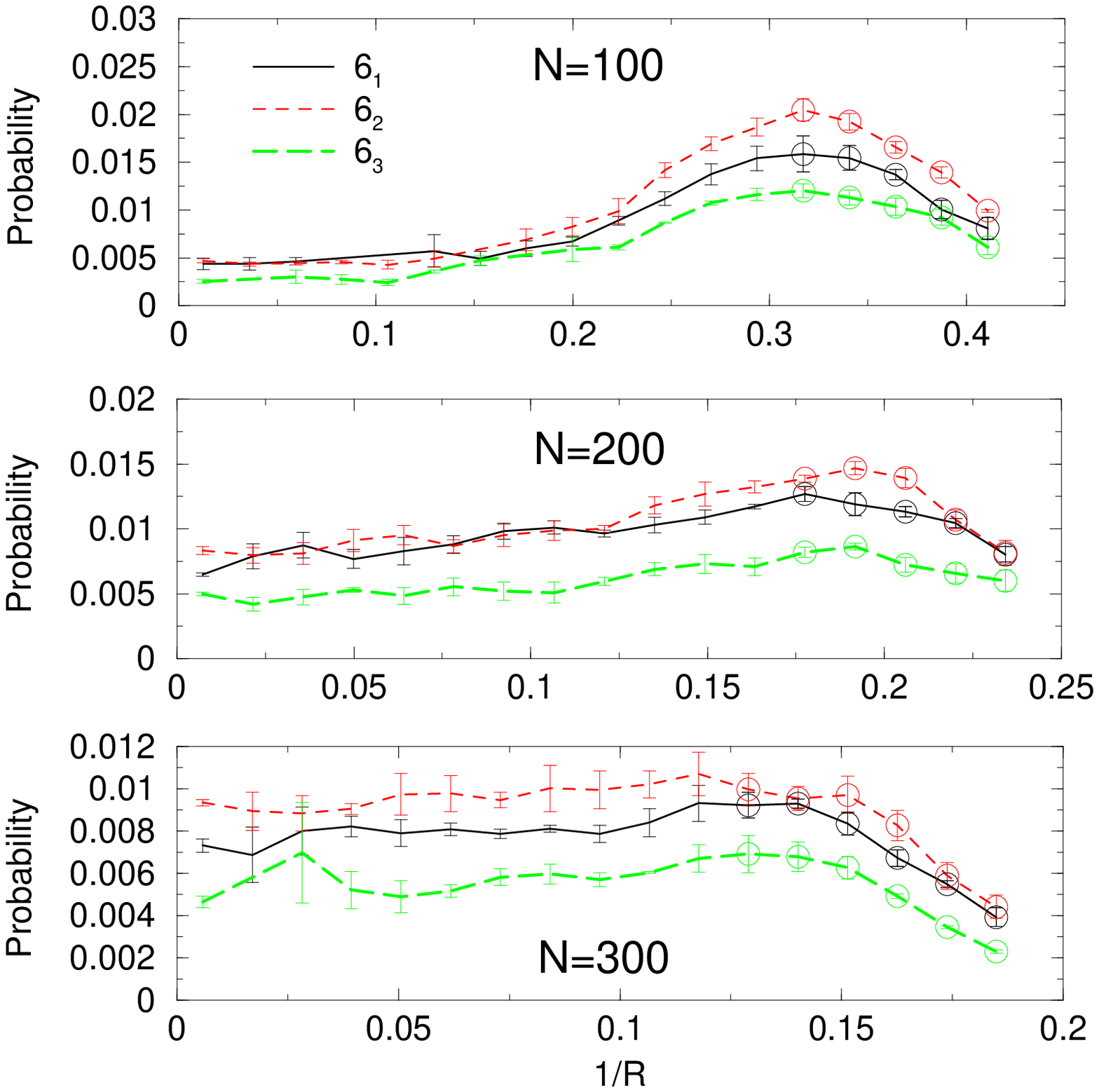}
\end{center}
\caption{}
\label{type6}
\end{figure}

\noindent Also the case of knots with $6$ minimal crossings is
particularly interesting since there are two chiral knots ($6_1$ and
$6_2$) and one achiral knot ($6_3$). The study of the relative
fraction of the chiral and achiral $6$ crossing knots shows
the effect of confinement on the
chirality of knots. Fig.~\ref{type6} portrays the
occurrence probability of $6_1$, $6_2$ and $6_3$ as a function of
$1/R$ for 3 different values of $N$. As one can see, the least
probable knot is the achiral one whose 
relative population among the 6 crossing knots,
within the explored
ranges of $N$,remains around 25\% as confinement is increased. This
illustrates that, in this simple case, confinement alone is not
sufficient to induce the chiral bias that is encountered in the
above-mentioned biological contexts.

In general, we find that the probability of formation of a given knot type
increases with $1/R$ -- with respect to its unconstrained value --
only if the chain length is small enough, specifically 
below the length that maximises the probability of formation 
of that knot type \cite{math_letters}.
As the confinement radius $R$ decreases we expect that the complexity
of the knots present in the ring would increase. This complexity can
be manifested either with the formation of prime knots with large
minimal crossing number or with the occurrence of composite knots,
i.e., knots that are connected sums of prime knots. It is
therefore interesting to monitor how the relative fraction of
composite/prime knots depends on $R$.

Figs.~\ref{knotprob2}f and \ref{knotprobs}f show the probabilities of
forming a composite knot~\cite{adams} for random chains of different
lengths. The trend follows the expectation that as confinement
increases the fraction of composite knots becomes higher.

\begin{figure}
\begin{center}
(a)\includegraphics[width=4.0in]{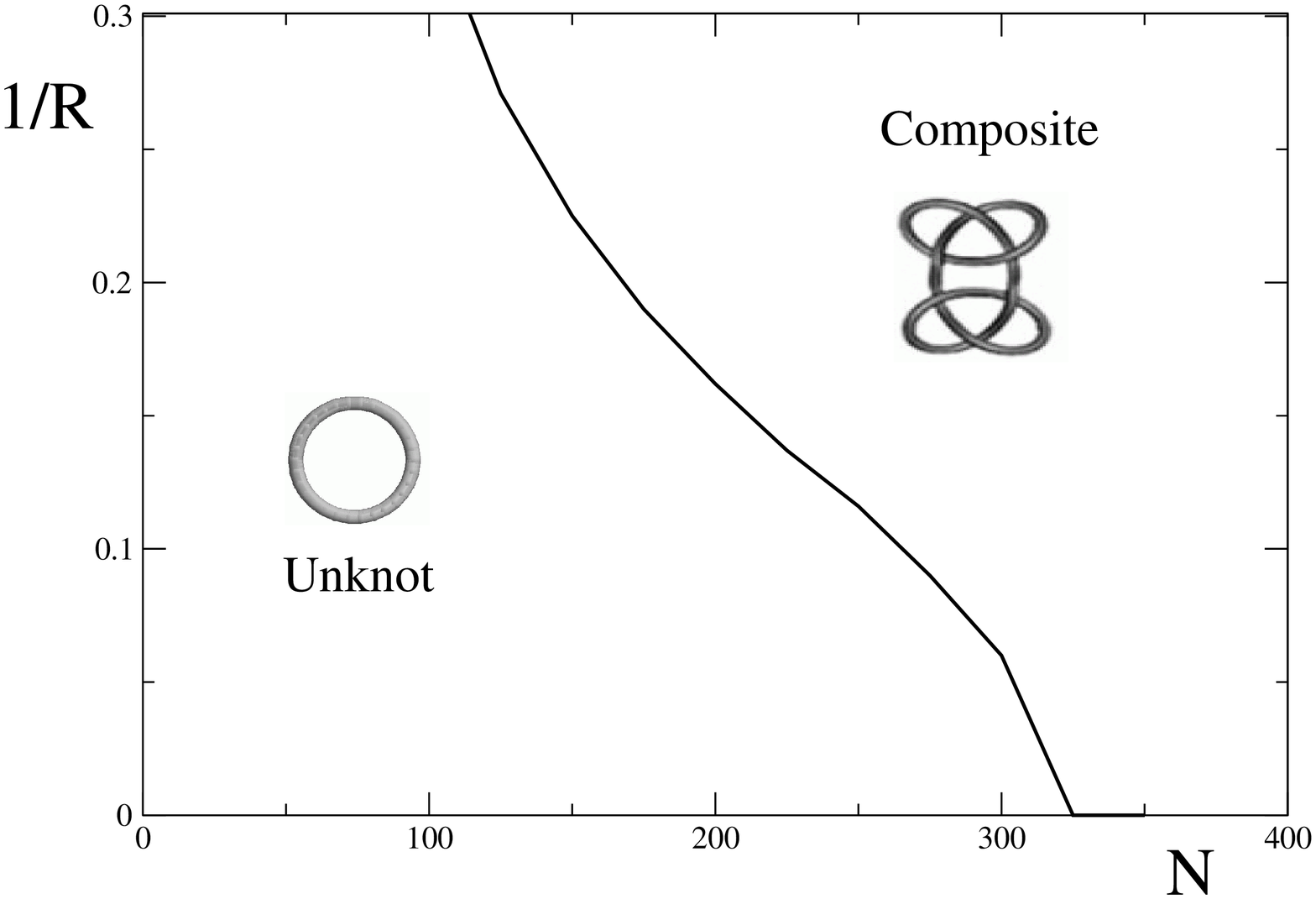}
(b)\includegraphics[width=4.0in]{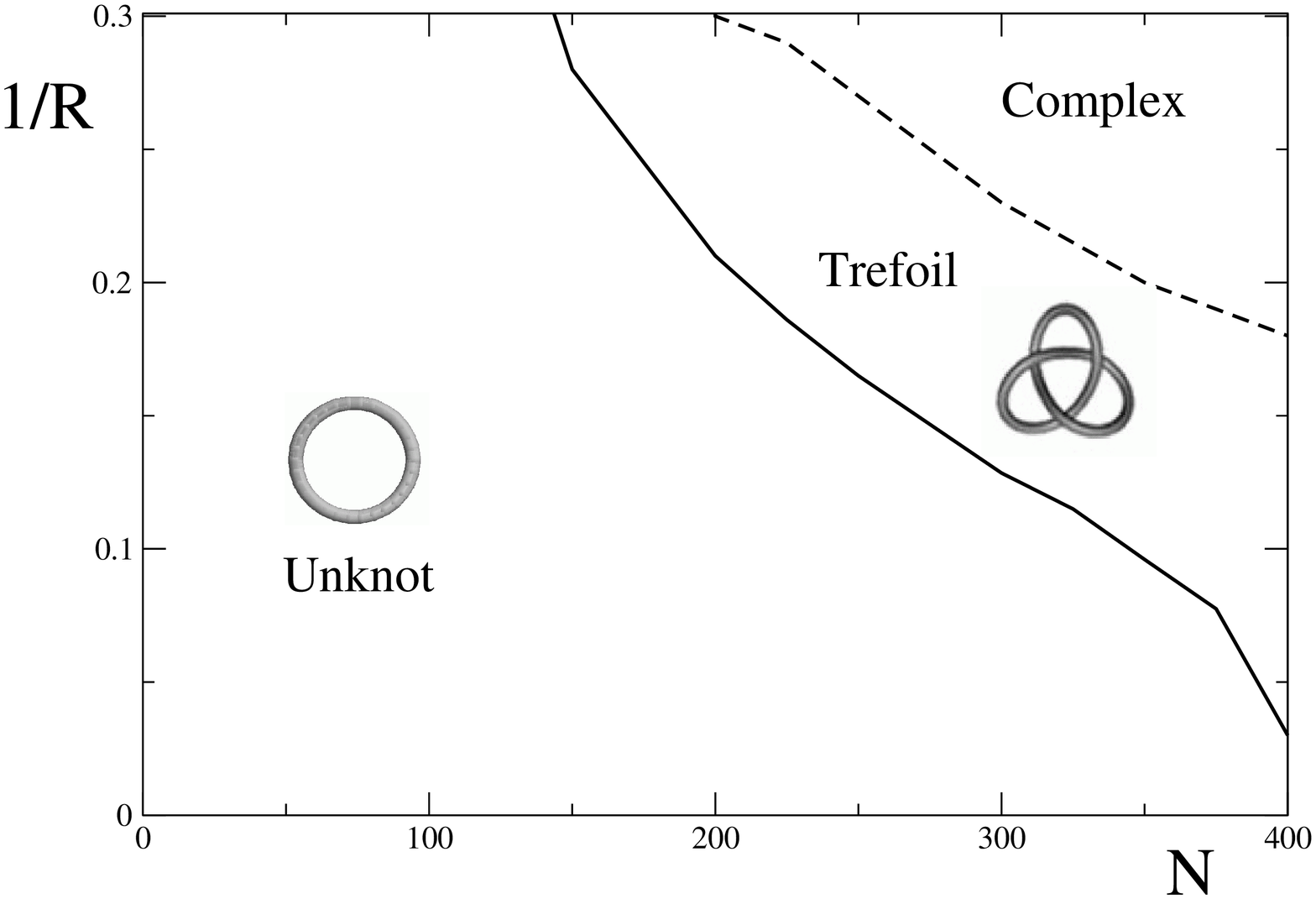}
\end{center}
\caption{}
\label{phase}
\end{figure}

Fig.~\ref{phase} shows a 'phase diagram' showing what knot class is
most populated. Throughout the values of the parameters
$\left(N,1/R\right)$ considered here, the most populated class was
either the unknot or the composite knots. As $N$ and $1/R$ increase,
i.e. as one moves to the right and to the top in the parameter space
in Fig.~\ref{phase}, knots become increasingly complex and prime knots
occur less and less often. Fig. \ref{phase}b shows the phase diagram
restricted to the ensemble of prime knots alone. 

\section{Writhe}

\begin{figure}
\begin{center}
\vskip 7.truecm
\includegraphics[width=4.0in]{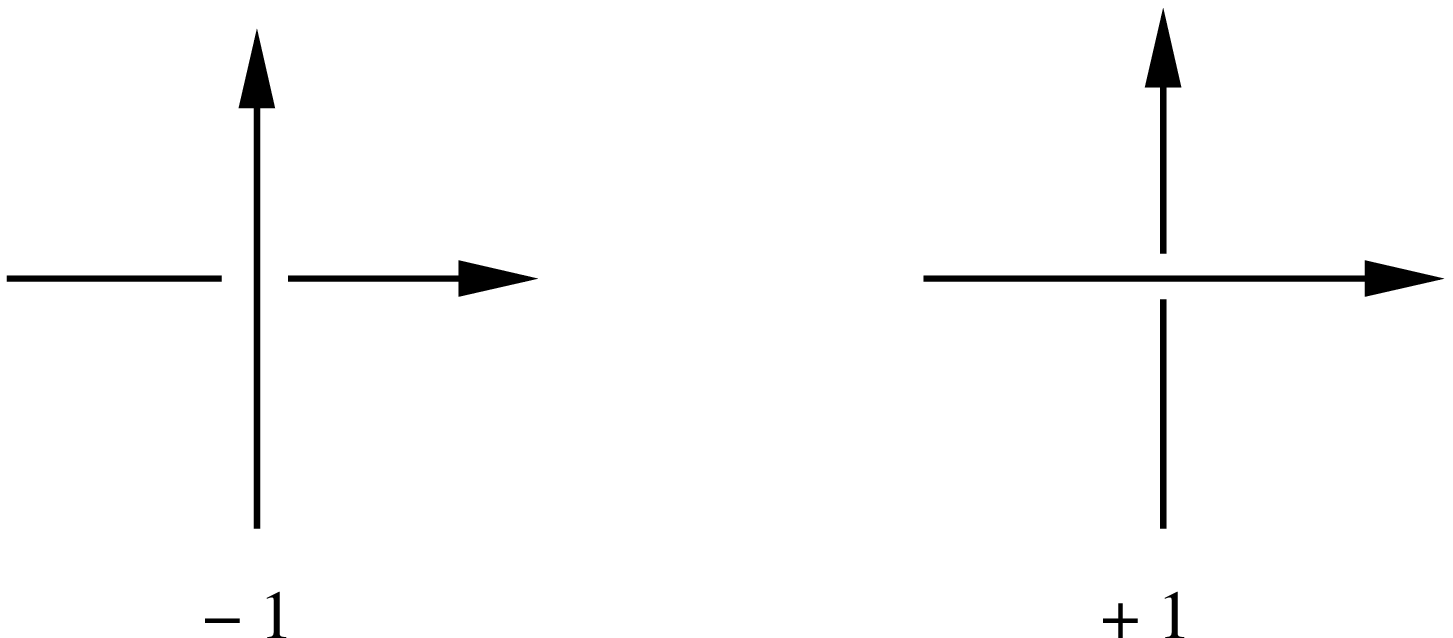}
\end{center}
\caption{}
\label{sign}
\end{figure}

Up to now we have focused on the topological properties of the system
but it also useful to have geometric measures of the polymer
entanglement. One interesting geometric property is the {\it writhe}
of the polymer, which has been proved to be useful in modelling the
degree of supercoiling in DNA\cite{Bauer80,White86}.  To define writhe
consider any oriented simple closed curve in ${\cal R}^3$, and project
it onto ${\cal R}^2$ in some direction $\hat{x}$. In general, the
projection will have crossings and, for almost all projection
directions, these crossings will be transverse, so that we can
associate a sign $+1$ or $-1$ to each crossing as in Fig.~\ref{sign}. 
The sign of a crossing is
independent of curve orientation, because changing the curve
orientation changes the orientation of both segments involved in a
crossing in a projection. For this projection we form the sum of these
signed crossing numbers, $S(\hat{x})$, and then average over all
projection directions $\hat{x}$. This average quantity is the writhe
$wr$ of the curve\cite{Fuller71}. If we compute the writhe of each
configuration with $N$ segments, and average over the set of
configurations, clearly the expected value of the writhe, $\langle wr
\rangle$, is zero by symmetry. Consequently, we shall be interested in
the expectation of the absolute value of the writhe, $\langle | wr |
\rangle$, or the expectation of its square, $\langle wr ^2 \rangle$
(or, more generally, in the distribution of $wr$).  The primary
difficulty with the computation of the writhe is that it involves
averaging the sum of signed crossing numbers over all projection
directions.  For polygons on the cubic lattice the calculation of
writhe is greatly simplified by a theorem\cite{Lacher91,Buks97} which reduces
the writhe computation to the average of linking numbers of the given
curve with four selected push-offs. Unfortunately this result is not
applicable here and we have to rely on the natural definition of the writhe
that we estimate by averaging $S(\hat{x})$ over more than 500 random
projections $(\hat{x})$.  Note that the writhe distribution before
and after the smoothing differ considerably (as should be expected and
consistently with lattice calculations \cite{enzo2}). The data on the
writhe presented here refer to the situation before
smoothing.

\noindent The reweighting technique can be used to
calculate the fraction
of conformations of given writhe, $wr$, that can
fit inside a sphere of radius, $R$, $W_N(wr,R)$. By necessity this can
be accomplished only after introducing a discretization of the values
for the writhe. The reweighting method therefore provides directly the
fundamental quantity of our interest, that is the probability
distribution for the writhe of rings of given lengths that fit in
hulls of given radius. From this fundamental quantity all the moments
of the distribution, including the averages $\langle wr \rangle$ and
$\langle wr^2 \rangle$ can be calculated:

\begin{equation}
\langle wr_N(R)^k  \rangle =
{ \int_{-\infty}^{+\infty}  dwr \ wr^k\, W_N(wr,R) \over 
\int_{-\infty}^{+\infty}  
dwr\ W_N(wr,R) }
\label{eqn:prob2}
\end{equation}

As far as the writhe is concerned we wish to elucidate two features
that have not been previously addressed in off-lattice contexts.  
The first one pertains to how, at given $N$, the probability distribution
for the writhe changes as rings are confined in tighter spaces while
the second concerns the $N$ dependence of the expectation of the
absolute value of the writhe for confined rings.

Fig.~\ref{writhe_invR_N} shows the (normalized) probability
distribution of the writhe for different values of $R$ and $N$. One
can notice that, for fixed $N$, as the confining sphere decreases, the
writhe distribution becomes broader and broader keeping their averages
equal to zero (as it should be since we are considering the whole set
of rings).  For any given $R$ the width of such distributions can
be measured, for example, by computing the mean of the absolute value
of the writhe $\langle | wr_N(R) | \rangle$. It is known that in the
unconstrained case, $R=\infty$, the writhe spread is proportional to
$\sqrt{N}$\cite{Janse93}. Accordingly, the ratio ${\langle
|wr_N(R=\infty)| \rangle}/ \sqrt{N}$ will be independent of $N$. To
check if this simple scaling behaviour holds also for finite values of
$R$, we have considered the quantity $\log[{\langle |wr_N(R)|
\rangle}/ \sqrt{N}]$, which is reported in Fig.~\ref{abswr_vs_R_scal}.
As expected, the curves take on the same value for $1/R \to 0$. This
collapse does not extend, however, for finite $R$'s. The tendency
towards linear behaviour that is visible for sufficiently large $1/R$
(i.e. for strong confinement) is suggestive of an exponential
dependence of the writhe on $1/R$ for all values of $N$.

The second feature we wish to elucidate is if and how the $N$
dependence of the absolute value of the writhe, is affected by the
confinement. For unconstrained polygons on the cubic lattice rigourous
arguments have shown that $ \langle |wr | \rangle$ cannot grow slower
than $\sqrt N$\cite{Janse93} and numerical results gives a power law
behavior $\langle |wr | \rangle \sim N^\alpha$ with $\alpha \sim
0.52\pm 0.04$\cite{Janse93}. To the best of our knowledge no such
investigations have been carried out for the case of confined polymer
rings, though it has been argued that for highly compact conformations
of a simple closed curve the writhe increases like
$N^{4/3}$\cite{Cantarella97}. In Fig \ref{fig:alpha} a log-log plot of
$\langle |wr | \rangle$ vs $N$ is reported for the unconstrained case
(lower curve) and for rings of different length but confined in a
sphere with the same radius. Both curves appear to follow a power law
behavior. A linear regression in the log-log plot gives an exponent
equal to $0.498\pm 0.002$ for the $R=\infty$ case, in excellent
agreement with what is expected for unconstrained rings. For the
particular choice of confining radius of Fig. \ref{fig:alpha},
instead, the exponent is approximately 0.75, strikingly different from
the square root behavior. Indeed, the deviations from the $R=\infty$
behaviour appear to increase as a function of confinement. This is in
accord with the intuitive expectation that confinement can
dramatically increases the geometrical compexity of the rings, and
hence impact on their writhe. The linear trend of the finite-$R$ data
of Fig. \ref{fig:alpha} is suggestive of the existence of
$R$-dependent scaling laws for $\langle |wr_N(R)| \rangle$ as a
function of $N$. To reach a definite conclusion about the existence of
such peculiar scaling behaviour it would be necessary to extend
significantly the range of values of $N$ to be explored.

\section{Discussion and conclusions}

We investigated the occurrence of various types of knots in random
rings subject to spatial confinement. The sampling of ring
conformations was carried out within a multiple Markov chain
strategy. The correct identification of knotted conformations was
aided by a hierarchical simplification of the ring conformations by
first smoothing and shrinking the rings and then applying the
operations and calculations of KnotFind to their two-dimensional
projections.

The fraction of composite knots is shown to increase significantly with
both their length, $N$, and the inverse radius of the confining
sphere, $1/R$. Indeed, in tightly confined geometries the majority of
knots are composite. Furthermore, the probability of occurrence for
the simplest prime knots displays a non-monotonic behaviour: the
initial increase with confinement is followed by a subsequent decay at
still higher compression. This trend is similar to that observed for
unconstrained random knots for increasing length.

The scaling behaviour of the unknotting probability was also
investigated. Owing to the powerful numerical strategy adopted here,
we improve a previous estimate of the exponent governing the scaling
dependence on $N$ and $1/R$. Finally, the analysis of the writhe
distribution at fixed ring length suggests an exponential broadening
of the distribution under strong confinement. Also, if $N$ is
increased at fixed $R$ it is found that the writhe distribution width
increases more rapidly than in the case of unconfined knots.

Despite the fact that the rings considered here are not subject to
volume exclusion interaction it is pleasing that the observed increase
of the knotting probability with confinement is in line with the
experimental findings on DNA \cite{arsuaga,dewit2}. It is our
intention, for the future, to go beyond this significant qualitative
agreement and consider the more realistic case of rings with volume
exclusion. Indeed previous results \cite{cozzarelli,kamenetski}
suggest that without confinement ($R=\infty$), for a fixed $N$ the
knot probability is very sensitive to volume exclusion -- as one
increases the volume of the segments, the knot probability decreases
very quickly. The algorithm employed here ought to be efficient also
for the sampling of confined rings with excluded volume, though we can
envisage that the increased computational complexity will prevent us
from attaining the same ring lengths or degree of confinement obtained
here.

This work was funded by INFM. DWS was partially supported by a
Burroughs Wellcome Fund Interfaces grant to the Program in Mathematics
and Molecular Biology. CM and DWS acknowledge the hospitality of the
Institut des Hautes Etudes Scientifiques in Bures-sur-Yvette where
part of this study was undertaken. We are grateful to Ken Millet, Eric
Rawdon, Andrzej Stasiak and Tommaso Zillio for useful discussions and
comments.

\newpage

\begin{center}
{\bf Figure Captions}
\end{center}

{{\bf Figure 1:}Example of configurations at four stages of the smoothing
procedure of two rings of 100 links. a-d refer to an unknot while e-h
to a trefoil knot.} \\

{{\bf Figure 2:} Probability of unknown knots as a function of $N$.} \\

{{\bf Figure 3:} Length dependence of the probability of occurrence of various
 types of knots in unconstrained random rings ($R=\infty$, the number
of crossing is indicated in the legend). The dashed
 line is the single exponential fit through the unknot data, yielding
 a decay exponent equal to $b= 0.0041\pm 0.0001$ (see text).} \\ 

{{\bf Figure 4:} Probabilities of formation of various types of knots of $N
=\{ 100, 200, 300, 400\}$ segments as a function of the inverse radius
of the enclosing sphere, $1/R$. Open symbols denote the region when
the fraction of unknown knots exceeds 10\% (but is smaller than
50\%). No data is plotted in the region when the fraction of unknown
knots exceeded 50\%. The plots refer to knots of type: (a) unknot, (b)
$3_1$, (c) $4_1$, (d) $5_1$ and $5_2$, (e) $6_1$, $6_2$ and $6_3$ and
(f) composite.} \\

{{\bf Figure 5:} Probabilities of formation of various types of knots as a
function of the ring length $N$ and inverse radius of the enclosing
sphere, $1/R$. At values of $1/R$ greater than those indicated with
the red line, the observed fraction of unknown knots exceeded 10\%
(but is smaller than 50\%). No data is plotted in the region when the
fraction of unknown knots exceeded 50\%. The plots refer to knots of
type: (a) unknot, (b) $3_1$, (c) $4_1$, (d) $5_1$ and $5_2$, (e)
$6_1$, $6_2$ and $6_3$ and (f) composite.} \\

{{\bf Figure 6:} Scaling of the unknotting probabilities for a random rings
with number of segments ranging from 100 to 450 in steps of 25.} \\

{{\bf Figure 7:} Scaling of the trefoil probabilities for a random 
chain with $N$ from 250 to 450 in steps of 25.} \\

{{\bf Figure 8:} The plot shows the length of the rings, $N$, above which the
  $k=5$ knots outnumber the $K=4$ ones.}\\

{{\bf Figure 9:} Dependence on confining radius of the probability of forming
a $6_{1,2,3}$ knot of $N=\{100, 200, 300\}$ segments.}\\

{{\bf Figure 10:} (a) ``Phase diagram'' showing, for each point in the
$\left(N,1/R\right)$ plane considered in the simulations, what is the
most populated class of knots. In panel (b) the diagram is restricted
to the class of prime knots alone. The dashed line indicates the
threshold between trefoil and more complex knots, but it has been
computed in a region where the unclassified knots are the majority.}\\

{{\bf Figure 11:} Positive and negative crossings are determined by a 
right-hand rule.}\\

{{\bf Figure 12:} Probability distribution for the writhe. Each figure
corresponds to a fixed value of $N$ (top: $N=100$, midlle $N=200$ and
bottom $N=300$) and different curves correspond to different $R$
values.}\\

{{\bf Figure 13:} Dependence of the logarithm of $\langle |wr|
\rangle/\sqrt(N)$ on $1/R$. Different curves correspond to different
values of $N$.}\\

{{\bf Figure 14:} Log-log plot of the absolute value of the writhe as a
function of the ring length, $N$. The two data sets corresponds to the
unconstrained case (black circles) and to polymer rings confined in a
sphere of inverse radius $1/R=0.226$ (squares).} \\

\newpage

\newpage

\newpage

\newpage

\newpage

\newpage

\newpage

\newpage

\newpage

\newpage

\newpage

\newpage

\begin{figure}
\phantom{.}
\vskip 5.truecm
\begin{center}
\includegraphics[width=4.0in]{writhe_invR_N.eps}
\end{center}
\caption{}
\label{writhe_invR_N}
\end{figure}

\newpage

\begin{figure}
\phantom{.}
\vskip 5.truecm
\begin{center}
\includegraphics[width=4.0in]{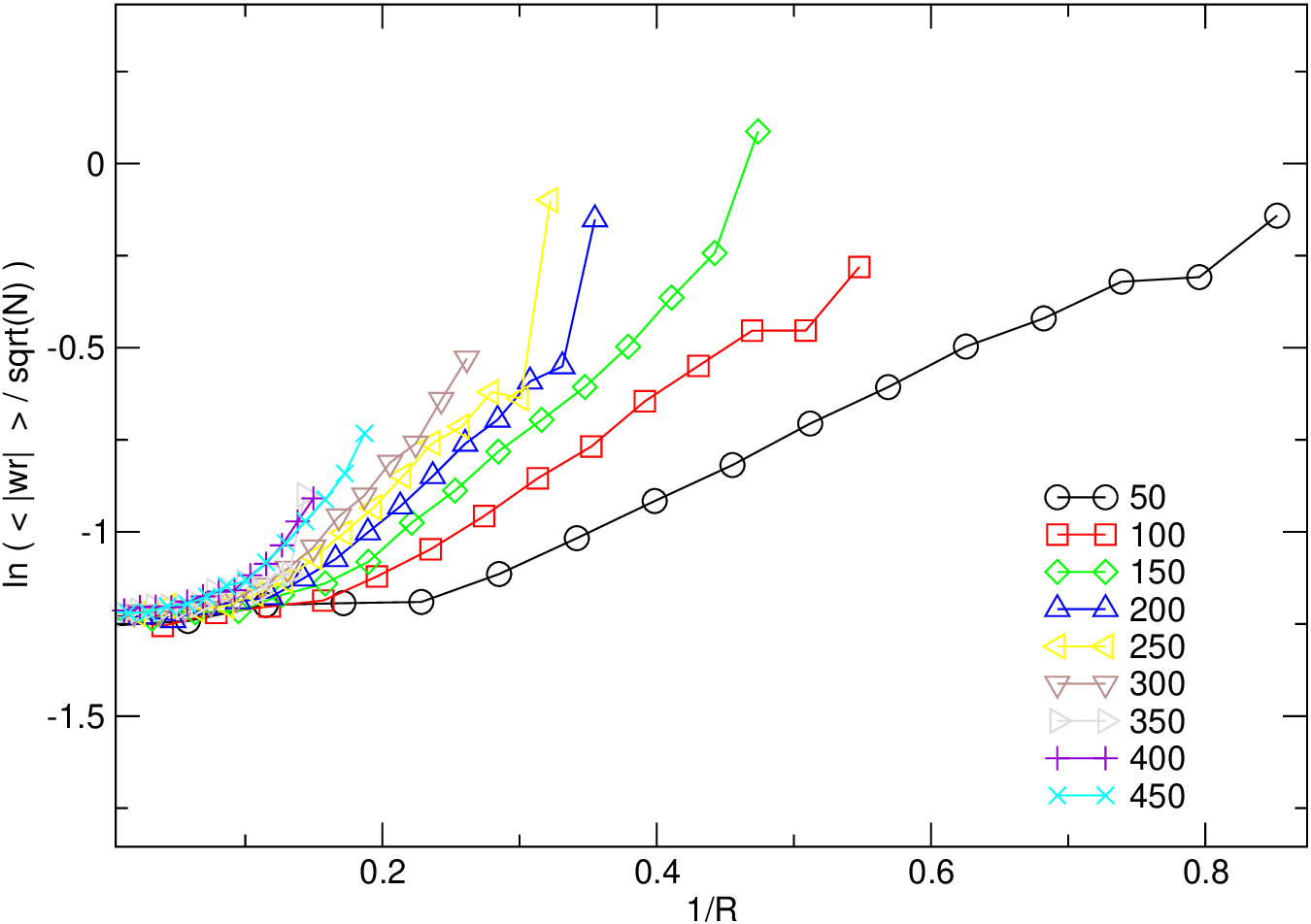}
\end{center}
\caption{}
\label{abswr_vs_R_scal}
\end{figure}

{}

\newpage

\begin{figure}
\begin{center}
\vskip 5.truecm
\includegraphics[width=4.0in]{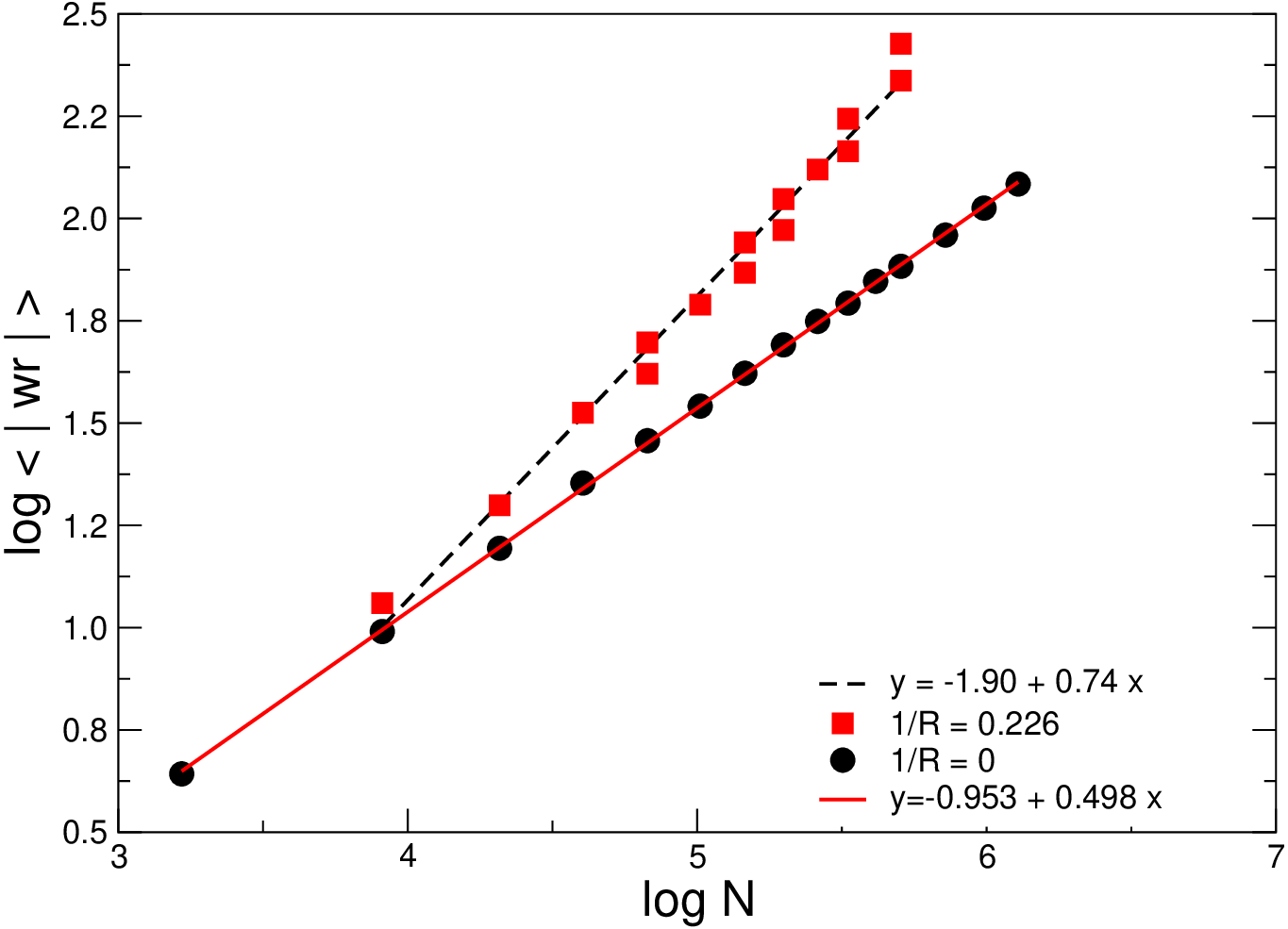}
\end{center}
\caption{}
\label{fig:alpha}
\end{figure}

\end{document}